\documentclass[iop, superscriptaddress, reprint, graphicx]{revtex4-1}
\usepackage{braket}

\usepackage{color}

\newcommand{\lp}{\left(}
\newcommand{\rp}{\right)}
\newcommand{\lab}{\left<}
\newcommand{\rab}{\right>}
\newcommand{\lsb}{\left[}
\newcommand{\rsb}{\right]}

\newcommand{\labs}{\left|}
\newcommand{\rabs}{\right|}

\newcommand{\refeq}[1]{Eq. (\ref{#1})}
\newcommand{\refsec}[1]{Sec. \ref{#1}}
\newcommand{\refapp}[1]{Appendix \ref{#1}}
\newcommand{\reffig}[1]{Figure \ref{#1}}

\newcommand{\abs}[1]{\labs #1 \rabs}

\newcommand{\KAISTNQe}{Department of Nuclear and Quantum Engineering, KAIST, Daejeon 34141, Korea}

\newcommand{\NFRI}{National Fusion Research Institute, Daejeon, Republic of Korea}
\newcommand{\Wigner}{Wigner RCP, Euratom Association-HAS, Budapest, Hungary}

\newcommand{\jwkim}{\author{Jaewook~Kim}\email[]{ijwkim@kaist.ac.kr}\affiliation{\KAISTNQe}}

\newcommand{\yunam}{\author{Y.U.~Nam}\affiliation{\NFRI}}
\newcommand{\mlampert}{\author{M.~Lampert}\affiliation{\Wigner}}

\newcommand{\ycghim}{\author{Y.-c.~Ghim}\email[]{ycghim@kaist.ac.kr}\affiliation{\KAISTNQe}}

\usepackage{amsmath}
\usepackage{graphicx}
\usepackage{epstopdf}

\draft 
\usepackage[english]{babel}

\newcommand{\DT}{\Delta T}
\newcommand{\DTsub}{\DT_\text{sub}}

\newcommand{\DL}{\Delta L}

\newcommand{\lx}{\lambda_x}
\newcommand{\tlife}{\tau_\text{life}}

\newcommand{\tac}{\tau_\text{ac}}

\newcommand{\varCorr}{\sigma^2_{_{C_{a,b}}}}
\newcommand{\varNormCorr}{\sigma^2_{_{\overline C_{a,b}}}}
\newcommand{\varNormCorrNoise}{\sigma^2_{_{\overline C_{a,b}^\text{noise}}}}
\newcommand{\NormCorrNoise}{\overline C_{a,b}^\text{noise}}
\newcommand{\elxM}{\ell_x}
\newcommand{\elxT}{\ell_x^\text{true}}

\begin{document}

\title{Reliability of the two-point measurement of the spatial correlation length from Gaussian-shaped fluctuating signals in fusion-grade plasmas}

\jwkim 
\yunam 
\mlampert 
\ycghim

\date{\today}
\begin{abstract}
A statistical method for the estimation of spatial correlation lengths of Gaussian-shaped fluctuating signals with two measurement points is examined to quantitatively evaluate its reliability (variance) and accuracy (bias error). The standard deviation of the correlation value is analytically derived for randomly distributed Gaussian shaped fluctuations satisfying stationarity and homogeneity, allowing us to evaluate, as a function of fluctuation-to-noise ratios, sizes of averaging time windows and ratios of the distance between the two measurement points to the true correlation length, the goodness of the two-point measurement for estimating the spatial correlation length. Analytic results are confirmed with numerically generated synthetic data and real experimental data obtained with the KSTAR beam emission spectroscopy diagnostic. Our results can be applied to Gaussian-shaped fluctuating signals where a correlation length must be measured with only two measurement points.
\end{abstract}

\maketitle

\section{introduction}
\label{sec:intro} 
Plasma turbulence is an intellectually interesting phenomenon as we do not have full capability of predicting a cause, an evolution and a consequence of it; yet it heavily influences the world around us. For instance, in both laboratory and astrophysical plasmas, a possible explanation for the amplification of magnetic fields with \textit{turbulence} is provided \cite{Meinecke_PNAS_2015} (and the references therein); while plasma \textit{turbulence} is believed to be one of the major obstacles for harnessing nuclear fusion power economically \cite{Carreras1997}. 

It is, therefore, important to acquire and provide statistically accurate (precision to the true value or bias error) and reliable (variance of the measured values) experimental measurements of turbulence. Probably, the best and easiest way to do so is by using a large number of temporally fast detectors covering the space of interest. Temporally fast detectors are becoming widely available; on the other hand, using a large number of such detectors may not be a feasible solution due to many reasons such as installation difficulties, bandwidth processing and limited resources on budgets depending on applications. Also, we do not want to cover the whole space just with the detectors if they are in-situ types. For this reason, there has been previous studies on obtaining the spatial structure of fluctuating signals with two measurement points in laboratory plasmas \cite{Iwama1979, Beall1982} and with four measurement points in solar winds \cite{horbury_clusterII_2000}. Note that this four-point measurement in solar winds is really a two-point measurement as the four points are not aligned in a straight line (or to a background magnetic field line). 

As we do not find any systematic studies on the reliability of the obtained two-point correlation function, i.e., variance of the correlation function, which is equally as important as the correlation function itself, we investigate how accurately and reliably one can measure the correlation function with only two measurement points. This is then used to calculate the accuracy and reliability of the estimated spatial correlation length. 

We focus on the \textit{two-point} measurement because it is the smallest necessary number to get the spatial structure unless Taylor's hypothesis \cite{Taylor_PRSL_1938} can be validly applied where temporal information at one spatial position contains upstream spatial information. We focus on the \textit{spatial correlation length} because it is one of the basic properties of turbulence. Decorrelation rate in time and fluctuation levels are also important characteristics, however these suffer less from system hardware due to availability of many time points. Furthermore, turbulence is believed to be critically balanced \cite{goldreich_apj_1995, cho_apj_2004, schekochihin_apjs_2009, nazarenko_jfm_2011} as reported in observations of solar winds \cite{horbury_prl_2008, podesta_apj_2009, wicks_mnras_2010} and a gyro-kinetic simulation of ion-temperature-gradient driven turbulence \cite{barnes_prl_2011} in a fusion relevant geometry. However, in fusion experiments the parallel correlation length has never been measured in the core of the plasmas, consequently only an experimental `signature' of critically balanced turbulence in MAST (Mega Amp Spherical Tokamak) is reported \cite{ghim_prl_2013}. The parallel correlation length of fluctuations are typically measured using the probes only at the edge of the fusion-grade plasmas \cite{Ritz_rsi_1988, Winslow_rsi_1997, Thomsen_pop_2002}. As the diagnostic systems being developed, there is a possibility of measuring parallel correlation lengths in the core of KSTAR (Korea Superconducting Tokamak Advanced Research) plasmas with the beam emission spectroscopy (BES) \cite{Lampert2015} and microwave imaging reflectometry (MIR) \cite{Lee_jinst_2012}. Both systems are 2D (poloidal and radial) but installed at different toroidal locations measuring the same physical quantity - density fluctuations. Again, this motivates us to study the reliability and accuracy of the correlation lengths obtained with the two measurement points.

This paper is organized as follows: in \refsec{sec:corr_fcn_eddy}, the correlation function and its variance of randomly distributed moving Gaussian-shaped (both in time and space) fluctuations are analytically derived with a finite averaging time window and noise. This expression is, then, compared with the numerically generated synthetic data where we find good agreement between the analytic and numerical results. This correlation function is used to obtain the accuracy (bias error) and reliability (variance) of the estimated correlation length in \refsec{sec:two_point_measurement} both analytically and numerically whose results are also confirmed quantitatively with the experimental data obtained from the KSTAR BES system. Our conclusions are provided in \refsec{sec:conclusion}.

\section{Correlation function of fluctuating signals and its variance}
\label{sec:corr_fcn_eddy}

We model a time ($t$) dependent 1D fluctuating signal at the spatial location $x=x_a$ as the sum of `eddies' :
\begin{equation}
\label{eq:eddy_sum}
S_{a}(t)={\displaystyle \sum\limits _{i=1}^{N}S_{a_{i}}(t)},
\end{equation}
where $N$ is the total number of eddies and $S_{a_{i}}(t)$ is the signal due to $i^\text{th}$ `eddy' at the location $x=x_a$. As has been done previously \cite{Balazs2011, Ghim2012, jwkim_cpc_2016}  motivated by the experimental observations \cite{Fonck1993, Mckee2003} of ion-scale turbulence in fusion-grade plasmas, we use a Gaussian-shaped structure in both time and space for each eddy:
\begin{eqnarray}
\label{eq:one_eddy}
S_{a_i}\lp t\rp = A_i \exp &&\lsb -\dfrac{\lp t-t_i\rp^2}{2\tlife^2}\right. \nonumber \\
&&\;~-\left.\dfrac{\lp x_a-v\lp t-t_i\rp - x_i \rp^2}{2\lx^2} \rsb,
\end{eqnarray}
where the coherent properties of each eddy in time and space are parameterized by the characteristic temporal scale ($\tlife$) and the spatial scale ($\lx$). Furthermore, we allow the eddy to move with the speed of $v$ to mimic observed eddy motions due to background plasma flows \cite{Ghim2012} or wave-like propagations. We let the $i^\text{th}$ eddy have the maximum amplitude $A_i$ at $x=x_i$ and $t=t_i$ where random number $A_i$ is selected from a normal distribution with zero mean and variance of $A^2$. $x_i$ and $t_i$ are selected from uniformly distributed random numbers within the finite domain of $\lsb -\frac{\DL}{2}, \frac{\DL}{2}\rsb$ and $\lsb -\frac{\DT}{2}, \frac{\DT}{2}\rsb$, respectively. This means that our result is strictly valid within a flux-surface where the radial gradients of various equilibrium quantities correlated with the turbulence \cite{Ghim_nf_2014} are constant.

A spatio-temporal filling factor $F$ is defined as \cite{Ghim2012}
\begin{equation}
\label{eq:filling_factor}
F\equiv N\lp{\displaystyle \frac{\lx}{\DL}}\rp\lp{\displaystyle \frac{\tlife}{\DT}}\rp,
\end{equation}
and we control the total number of eddies $N$ such that $F\sim\mathcal{O}\lp 1\rp$ not to have too frequent or too rare eddies. Later on, we show that 
 $F\sim\mathcal{O}\lp 1\rp$ ensures the generated signal to have the square of the fluctuation level indeed of the order of $A^2$ as specified. More detailed descriptions on the model of the fluctuating signal can be found elsewhere \cite{Balazs2011, jwkim_cpc_2016}.

For the readers who question the validity of the Gaussian-shaped eddies as the Lorentzian eddies are also observed in the scrape-off-layer region of the magnetically confined plasmas \cite{Maggs_PRL_2011, Hornung_PoP_2011, DIppolito_PoP_2011}, we provide the discussions with the Lorentzian eddies in \refapp{app:Lorentzian}. This section is recommended to be read after reading \refsec{sec:two_point_measurement} .

\subsection{Correlation function and its variance}
\label{sec:analytic_expression}

In this section, we analytically derive the correlation value between the two spatial positions at $x=x_a$ and $x=x_b$ at the correlation time delay $\tau=0$ (as we are interested in obtaining the correlation length of the fluctuating signals) and its associated variance.

To analytically derive the correlation function following Kim et al. \cite{ jwkim_cpc_2016}, $\DL$ and $\DT$ are assumed to be infinitely large compared to $\lx$ and $\tlife$. A correlation value $C_{a, b}$ between $x=x_a$ and $x_b$ is calculated as $\text{E}\lsb S_a\lp t\rp S_b\lp t\rp\rsb$, where $\text{E}\lsb \: \rsb$ is the time averaging operator over the `sub-time window' whose size is set by $\DTsub$. Then, the ensemble averaged correlation value $\lab C_{a,b}\rab$ is \cite{ jwkim_cpc_2016}
\begin{eqnarray}
\label{eq:corr_anal_final}
\lab C_{a,b} \rab & \approx & \pi A^2 F \lp\exp\lsb -\frac{\lp x_a -x_b\rp^2}{2\lp\sqrt 2\lx\rp^2}\rsb \right. \nonumber \\
&& \left. -2\sqrt\pi\frac{\tac}{\DTsub}\exp \lsb-\frac{\lp x_a-x_b\rp^2}{4\lp\lx^2+\tlife^2 v^2 \rp} \rsb\rp, \nonumber \\ 
& \approx & \pi A^2 F \lp\exp\lsb -\frac{\lp x_a -x_b\rp^2}{2\lp\sqrt 2\lx\rp^2}\rsb \rp,
\end{eqnarray} 
where the approximation in the last line is valid for $\tac/\DTsub\ll 1$. $\DTsub$ is much smaller than $\DT$ allowing many of $\DTsub$ to exist within $\DT$ (for an ensemble average \cite{jwkim_cpc_2016}) but still much larger than $\tac$, the usual auto-correlation time of the fluctuating signal in the lab frame defined as \cite{Bencze2005}
\begin{equation}
\label{eq:tauc}
\tac \equiv \dfrac{\lx\tlife}{\sqrt{\lx^2 + \tlife^2 v^2}}.
\end{equation}
Note that the square of the fluctuation level $\lab C_{a,a}\rab$ is $\sim\mathcal{O}\lp A^2\rp$ for the spatio-temporal filling factor $F\sim\mathcal{O}\lp 1\rp$.

The correlation values, i.e., $C_{a, b}$, have a certain distribution resulting from the fact that the amplitude ($A_i$) and the spatio-temporal location ($x_i$ and $t_i$) of the eddies are randomly distributed. This distribution gives us randomness in $C_{a, b}$, i.e., the variance $\varCorr$ (see \refapp{app:var_corr_value}), which is
\begin{eqnarray}
\label{eq:var_of_corr_approx}
\varCorr &\approx& \sqrt{2} \pi^{5/2} A^4 F^2 \dfrac{\tac}{\DTsub} \left( 1+ \exp\lsb -\dfrac{(x_a-x_b)^2}{2\lx^2} \rsb \right) \nonumber\\ 
&& + \dfrac{3}{\sqrt{2}} \pi^{3/2} A^4 F \dfrac{\tac}{\DTsub} \exp\lsb -\dfrac{(x_a-x_b)^2}{2\lx^2} \rsb,
\end{eqnarray}
where the approximation is valid for $\DL\gg\lx$, $\DT\gg\DTsub\gg\tlife\text{ (or }\tac\text{)}$, and $N\gg 1$ with the condition of spatio-temporal filling factor $F\sim\mathcal{O}\lp 1\rp$ consistent with the assumptions we made to obtain $\lab C_{a,b} \rab$ in \refeq{eq:corr_anal_final}. Note that large $N$ is, in fact, a consequence of having $F\sim\mathcal{O}\lp 1\rp$ with large $\DL$ and $\DT$.

\subsection{Normalized correlation function and its variance with and without noise}
\label{sec:analytic_expression_norm}

We now calculate the \textit{normalized} correlation functions with and without noise and their associated variances. Again, we are interested in the correlation value between the two spatial points $x_a$ and $x_b$ at the correlation time delay $\tau=0$. Although the correlation length can be estimated with the ensemble averaged correlation value derived above, we derive the normalized correlation value because the normalized one estimates the correlation length more accurately (smaller bias error) and reliably (smaller variance) compared to the unnormalized one, which we show later in \refsec{sec:unnormalized_vs_normalized}.

The ensemble averaged normalized correlation value without noise, $\lab\overline C_{a,b}\rab$, is
\begin{eqnarray}
\label{eq:corr_anal_simple}
\lab\overline C_{a,b}\rab&\equiv&\lab \dfrac{C_{a,b}}{\delta_a \delta_b} \rab \approx \lab \dfrac{C_{a,b}}{\delta_f^2} \rab \\ \nonumber
&\approx& \dfrac{\lab C_{a,b}\rab}{\lab \delta_f^2 \rab} =\exp\lsb -\frac{\lp x_a -x_b\rp^2}{2\lp\sqrt 2\lx\rp^2}\rsb,
\end{eqnarray}
where $\delta_a=\sqrt{C_{a,a}}$ and $\delta_b=\sqrt{C_{b,b}}$ are the fluctuation levels at the locations $x=x_a$ and $x_b$, respectively. We have denoted this fluctuation level as $\delta_f$ and assumed $\delta_a\approx\delta_f$ and $\delta_b\approx\delta_f$ based on the homogeneity of the data. Note that $\lab\delta_f^2\rab=\pi A^2F$ from \refeq{eq:corr_anal_final}. The validity of the approximation in the second line of \refeq{eq:corr_anal_simple} is provided in \refapp{app:ensemble_average_approx}.

As for the case of $C_{a,b}$, $\overline C_{a,b}$ also has a distribution resulting in a finite variance solely due to the randomness of the eddies. This variance denoted as $\varNormCorr$ is
\begin{equation}
\label{eq:var_of_norm_corr}
\varNormCorr\approx \dfrac{\varCorr}{\pi^2 A^4 F^2}\lp 1-\exp\lsb -\dfrac{(x_a-x_b)^2}{2\lx^2}  \rsb \rp^2.
\end{equation}
The detailed derivation of $\varNormCorr$ and assumptions we made are provided in \refapp{app:var_norm_corr_value}.

Let us now consider the signal with uncorrelated noise. Our model signal with noise $S_a^\text{noise}\lp t\rp$ is, then, 
\begin{equation}
\label{eq:signal_with_noise}
S_a^\text{noise}\lp t\rp = S_a\lp t\rp + n_a \lp t\rp,
\end{equation}
where $n_a\lp t\rp$ is the noise as a function of time at $x=x_a$. A normalized correlation value with noise ${\NormCorrNoise}^\star$ between the signals from two spatial positions $x_a$ and $x_b$ is
\begin{eqnarray}
{\NormCorrNoise}^\star=\dfrac{  \text{E} \lsb S_a(t) S_b(t) \rsb + \text{E} \lsb n_a(t) n_b(t) \rsb }{\sqrt{\text{E} \lsb S_a^2(t) \rsb + \text{E} \lsb n_a^2(t) \rsb}\sqrt{\text{E} \lsb S_b^2(t) \rsb + \text{E} \lsb n_b^2(t) \rsb}}. \nonumber \\
\end{eqnarray}
Note that $\text{E} \lsb n_a(t) n_b(t) \rsb=0$ unless $x_a=x_b$ due to the uncorrelated noise assumption. The normalized correlation value at $x_a=x_b$ is by definition one whether or not there exists noise, i.e., $\overline C_{a,a}=1$ and ${\overline C_{a,a}^\text{noise}}^\star=1$. It is obvious that the effect of $\text{E} \lsb n_a(t) n_b(t) \rsb$ on ${\NormCorrNoise}^\star$ for $x_a=x_b$ can be removed for our model signal, but it is also possible to do so with experimental data by considering the auto-correlation function as a function of the correlation time delay $\tau$ \cite{Zoletnik_PPCF_1998}. The idea is, basically, interpolating the auto-correlation signal around $\tau=0$ to eliminate the effect of $\text{E} \lsb n_a(t) n_b(t) \rsb$ present only around very short interval of $\tau=0$.

Once we remove the noise effect at $x_a=x_b$ both for our model data and experimental data, we can define $\NormCorrNoise$ as
\begin{eqnarray}
\NormCorrNoise&=&\dfrac{  \text{E} \lsb S_a(t) S_b(t) \rsb}{\sqrt{\text{E} \lsb S_a^2(t) \rsb + \text{E} \lsb n_a^2(t) \rsb}\sqrt{\text{E} \lsb S_b^2(t) \rsb + \text{E} \lsb n_b^2(t) \rsb}} \nonumber \\
&=&\dfrac{\delta_f^2}{\delta_f^2+\delta_n^2}\dfrac{C_{a,b}}{\delta_f^2},
\end{eqnarray}
where we assumed the the noise level $\delta_n$ is homogeneous as is the case for $\delta_f$. Then, the ensemble averaged normalized correlation value with noise is
\begin{eqnarray}
\label{eq:norm_corr_noise}
\lab\NormCorrNoise\rab &=& \lab \dfrac{\delta_f^2}{\delta_f^2+\delta_n^2} \rab \lab \dfrac{C_{a,b}}{\delta_f^2} \rab \nonumber \\
&\approx&\dfrac{\lab\delta_f^2\rab}{\lab\delta_f^2+\delta_n^2\rab} \lab\overline C_{a,b}\rab\nonumber\\
&=& \dfrac{\lab\delta_f^2\rab}{\lab\delta_f^2\rab+\lab\delta_n^2\rab} \lab\overline C_{a,b}\rab\nonumber\\
&=&\dfrac{\lab\delta_f^2\rab}{\lab\delta_f^2\rab+\lab\delta_n^2\rab}\exp\lsb -\frac{\lp x_a -x_b\rp^2}{2\lp\sqrt 2\lx\rp^2}\rsb.
\end{eqnarray}
The first line equality is owing to the fact the normalized correlation value $C_{a,b}/\delta_f^2$ is not correlated with a function of fluctuation levels, i.e., $\delta_f^2/\lp\delta_f^2+\delta_n^2\rp$. The second line approximation is based on the same rationale for $\lab \frac{C_{a,b}}{\delta_f^2} \rab \approx \frac{\lab C_{a,b}\rab}{\lab \delta_f^2 \rab}$ in \refeq{eq:corr_anal_simple} (see \refapp{app:ensemble_average_approx}). Notice that the effective role of the noise is to decrease the normalized correlation value, and it is not even unity at $x_a=x_b$, rather it is a function of fluctuation($\delta_f$)-to-noise($\delta_n$) ratio (FNR).

The variance $\varNormCorrNoise$ of the ensemble averaged normalized correlation value with noise calculated in \refapp{app:var_norm_corr_value_noise} is
\begin{eqnarray}
\label{eq:var_of_norm_corr_noise}
\varNormCorrNoise
&\approx&\lab\NormCorrNoise\rab^2 \nonumber \\
&&\lsb \lp\dfrac{\sigma_{\delta_f^2}}{\lab\delta_f^2\rab}-\dfrac{\sigma_{\delta_f^2}}{ \lab\delta_f^2\rab+\lab\delta_n^2\rab }\rp^2+\dfrac{ \varNormCorr }{ \lab\overline C_{a,b} \rab^2  }\rsb, \nonumber \\
\end{eqnarray}
where $\sigma^2_{\delta_f^2}$ and $\sigma^2_{\delta_n^2}$ are the variances of $\delta_f^2$ and $\delta_n^2$, respectively. Since $\delta_f^2=C_{a,a}$, we have $\sigma^2_{\delta_f^2}=\sigma^2_{_{C_{a,a}}}$ which can be calculated using \refeq{eq:var_of_corr_approx}. 

Note that \refeq{eq:norm_corr_noise} and \refeq{eq:var_of_norm_corr_noise} which are the ensemble averaged normalized correlation value and its variance with noise, respectively, become \refeq{eq:corr_anal_simple} and \refeq{eq:var_of_norm_corr} for $\delta_n^2=0$, and none of these equations contain unknowns for our model fluctuation data.

\subsection{Comparisons between analytic expressions and numerical results}
\label{sec:compare_eq_syn}

To confirm that the analytically derived unnormalized and normalized correlation values and their associated variances are indeed correct, we generate 50 sets of 1D synthetic data based on \refeq{eq:eddy_sum} and \refeq{eq:one_eddy}. These sets of synthetic fluctuation data are generated such that they are guaranteed to be homogeneous and stationary \cite{jwkim_cpc_2016}.

The basic properties of each set of the synthetic data are: the characteristic spatial scale $\lx=0.3$ m, the characteristic temporal scale $\tlife=15~\mu$s, the velocity of the eddies $v=5,000$ m/s, the total time window for the data $\DT=15,000~\mu$s with the sub-time window $\DTsub=585~\mu$s (giving 25 sub-time windows) and the spatial domain size $\DL=2.7$ m. To make sure that the synthetic data are homogeneous, $\delta_f^2$ is kept to be constant in space as well as satisfying the criteria on the domain size according to reference \cite{jwkim_cpc_2016}.

\begin{figure}[t]
\includegraphics[width=\linewidth]{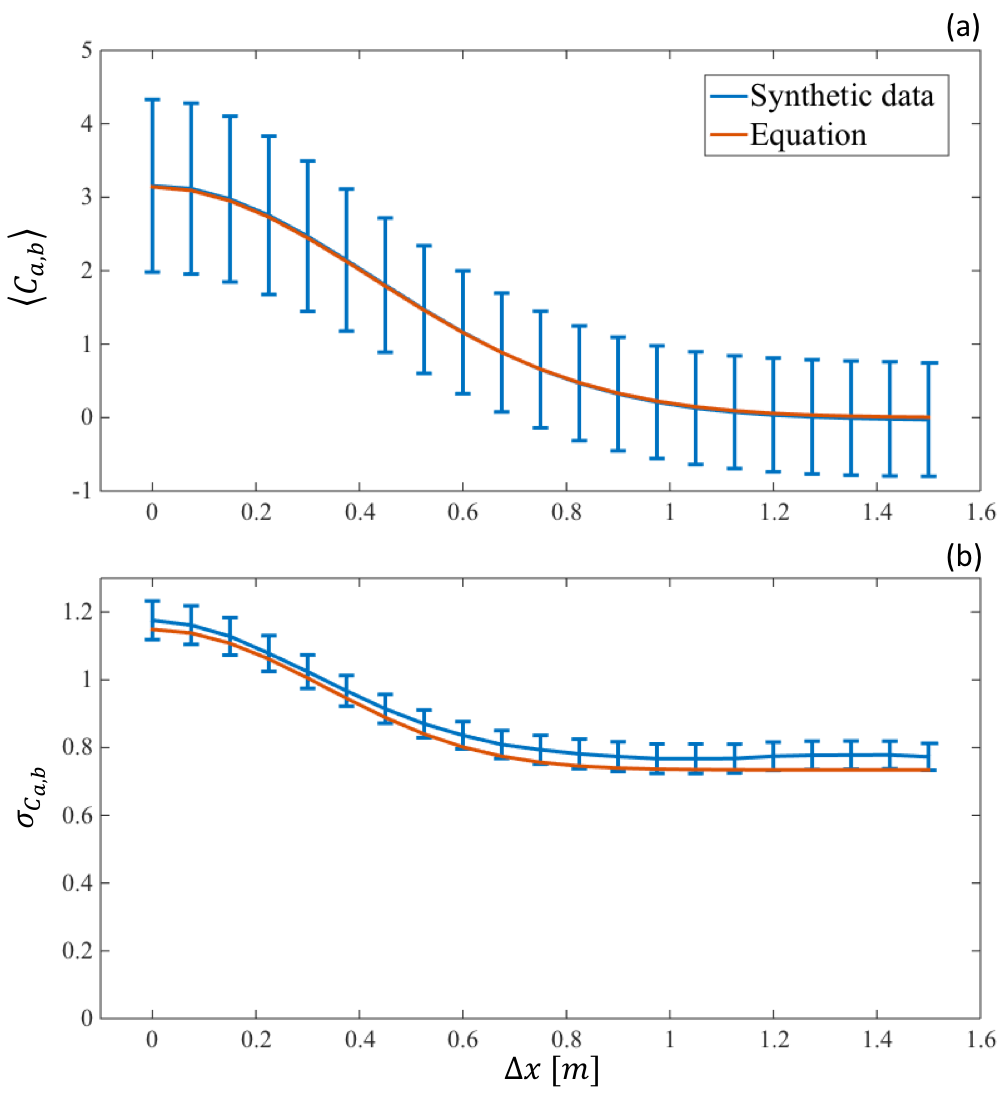}
\caption{(a) Correlation values and (b) their standard deviations as a function of $\Delta x=\abs{x_a-x_b}$ estimated from the synthetic data, $\lab C_{a,b}\rab^*$ and $\sigma_{_{C_{a,b}}}^*$ (blue lines), and calculated analytically, $\lab C_{a,b}\rab$ and $\sigma_{_{C_{a,b}}}$ (red lines). The error bars in (a) represent $\sigma_{_{C_{a,b}}}^*$, and those in (b) show $95$\% confidence interval.}
\label{fig:covariance_comparison}
\end{figure}

From the synthetic data we estimate the correlation value $\lab C_{a,b}\rab^*$ (we use a superscript asterisk to denote that it is estimated with the synthetic data)  and its standard deviation, i.e., square root of the variance, $\sigma_{_{C_{a,b}}}^*$ as
\begin{eqnarray}
\label{eq:syn_Cab_VarCab}
\lab C_{a,b} \rab^*= \dfrac{1}{50}\sum_{i=1}^{50} \underbrace{ \lsb \dfrac{1}{25}\sum_{j=1}^{25} C_{a,b}^{ij*} \rsb}_{\equiv \lab C_{a,b} \rab^{i*}}, \nonumber \\
\sigma_{_{C_{a,b}}}^* =  \dfrac{1}{50}\sum_{i=1}^{50}\underbrace{\lsb \sqrt{\dfrac{1}{25}\sum_{j=1}^{25} \lp C_{a,b}^{ij*}- \lab C_{a,b} \rab^{i*}\rp^2}  \rsb}_{\equiv \sigma^{i*}_{_{C_{a,b}}}}, 
\end{eqnarray}
where the superscripts $i$ and $j$ denote the $i^\text{th}$ set and the $j^\text{th}$ sub-time window within a set of the synthetic data, respectively. Here, $C_{a,b}^{ij*}$ is estimated as
\begin{equation}
\label{eq:Cabij_syn}
C_{a,b}^{ij*} = \dfrac{1}{\DTsub}\int_{\DTsub}S_a^{ij}\lp t\rp S_b^{ij}\lp t\rp dt.
\end{equation}

\reffig{fig:covariance_comparison}(a) shows analytically calculated $\lab C_{a,b}\rab$ (red line) using \refeq{eq:corr_anal_final} and $\lab C_{a,b}\rab^*$ (blue line) using \refeq{eq:syn_Cab_VarCab} on the synthetic data as a function of $\Delta x= \abs{x_a-x_b}$. They show a good agreement. The error bars on $\lab C_{a,b}\rab^*$ are $\sigma_{_{C_{a,b}}}^*$. Furthermore, \reffig{fig:covariance_comparison}(b) illustrates a good agreement within the $95$\% confidence interval between the analytically calculated $\sigma_{_{C_{a,b}}}$ (red line) using \refeq{eq:var_of_corr_approx} and estimated $\sigma_{_{C_{a,b}}}^*$ from the synthetic data using \refeq{eq:syn_Cab_VarCab} as a function of $\Delta x$. 

\begin{figure}[t]
\includegraphics[width=\linewidth]{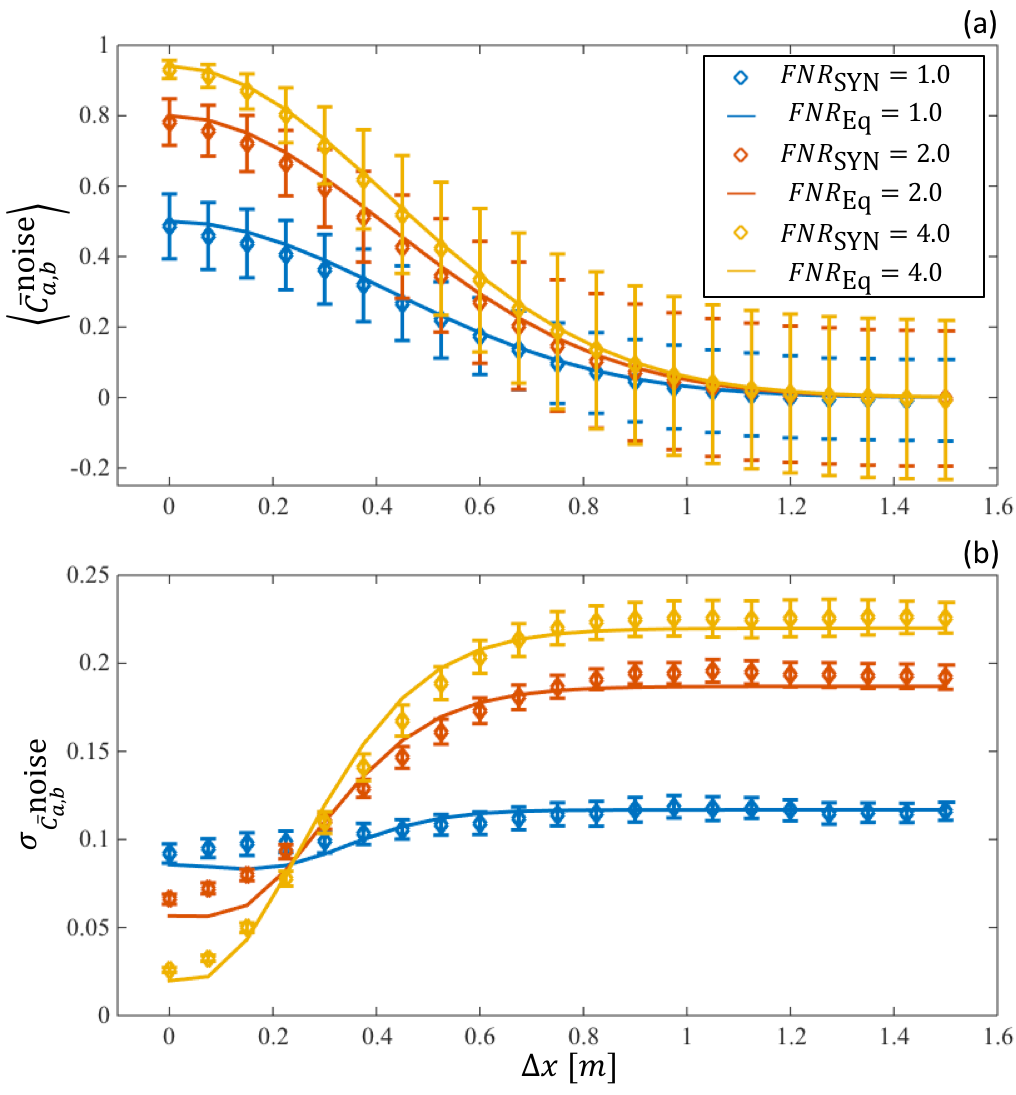}
\caption{Same as \reffig{fig:covariance_comparison} for (a) normalized correlation values and (b) their standard deviations estimated from the synthetic data (diamonds) and calculated analytically (lines) with the fluctuation-to-noise ratio (FNR) of $1.0$ (blue), $2.0$ (red) and $4.0$ (yellow).}
\label{fig:correlation_comparison}
\end{figure}

Similar comparisons are shown in \reffig{fig:correlation_comparison} for the normalized correlation values with various fluctuation-to-noise ratios (FNR=$\delta_f / \delta_n$): FNR = $1.0$ (blue), $2.0$ (red) and $4.0$ (yellow). Normalized correlation values and their standard deviations with noise, $\lab\NormCorrNoise\rab$ and $\sigma_{_{\overline C_{a,b}^\text{noise}}}$, are calculated using \refeq{eq:norm_corr_noise} and \refeq{eq:var_of_norm_corr_noise}; while they are also obtained from the synthetic data, $\lab\NormCorrNoise\rab^*$ and $\sigma^*_{_{\overline C_{a,b}^\text{noise}}}$, using \refeq{eq:syn_Cab_VarCab} with normalized $C_{a,b}^{ij*}$. Again, we find good agreements between the analytical and numerical results. 

Note that the value of $\lab\NormCorrNoise\rab$ with $x_a=x_b$, i.e., $\Delta x=0$ is $\lab \delta_f^2\rab/\lp\lab \delta_f^2\rab+\lab \delta_n^2\rab\rp$ according to \refeq{eq:norm_corr_noise}. Thus, we expect this value to be $0.50$, $0.80$ and $0.94$ for FNR = $1.0$, $2.0$ and $4.0$, respectively. These values coincide with the results from synthetic data which are obtained by removing the noise in the signal by using the interpolation technique around the correlation time delay $\tau=0$ of the auto-correlation function \cite{Zoletnik_PPCF_1998} as mentioned earlier.

As vindicated by the numerical results, we can use the analytically obtained normalized correlation values with and without noise and their associated variances to examine the accuracy (bias error) and reliability (variance) of the two-point measurement of the spatial correlation length.

\section{Accuracy and reliability of correlation length measurement with two spatial points}
\label{sec:two_point_measurement}

\begin{figure}[t]
\includegraphics[width=\linewidth]{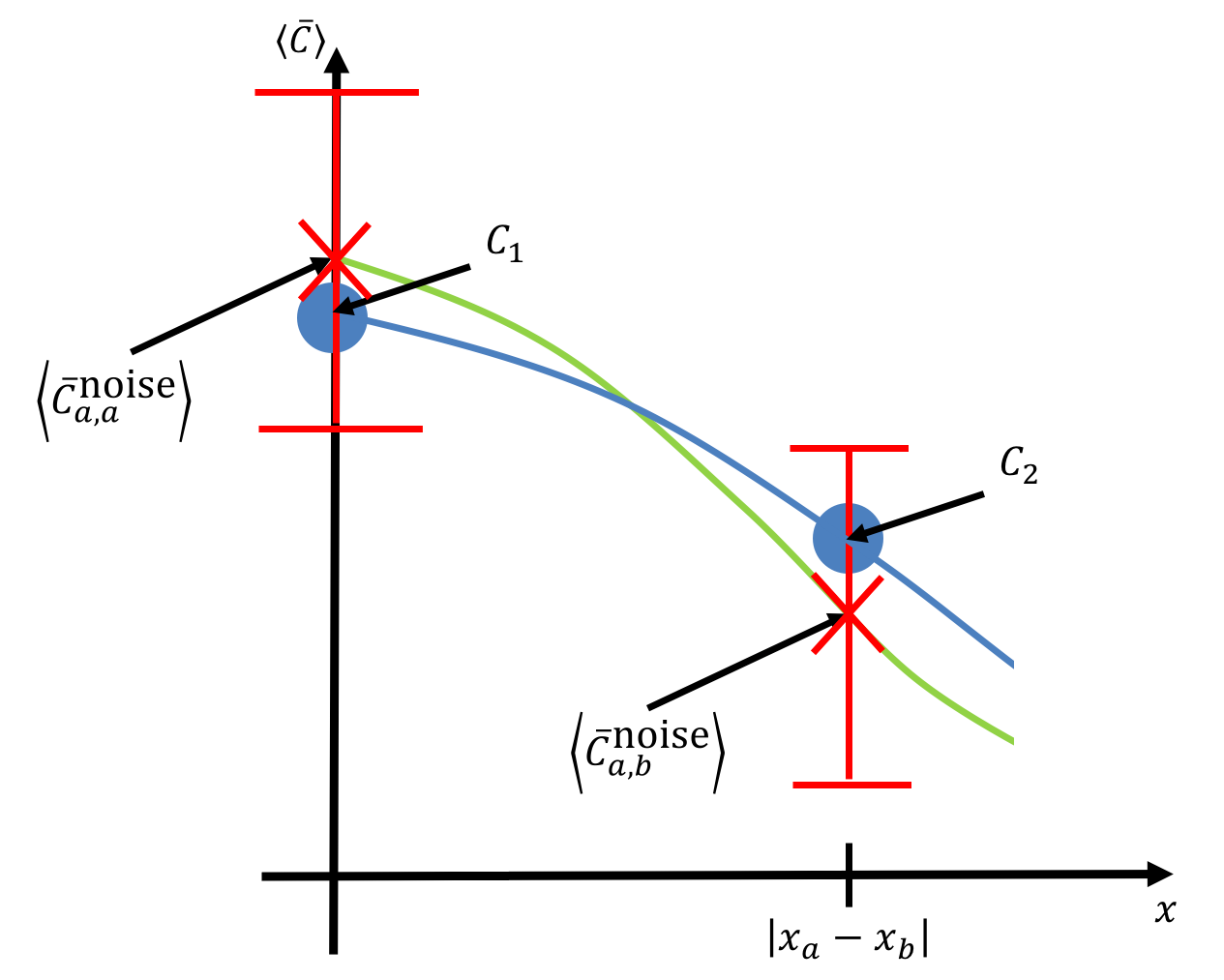}
\caption{A cartoon picture of the two points fitting. The normalized auto- and cross-correlation values $C_1$ and $C_2$ marked as blue filled circles, respectively, may be obtained experimentally; while the true values are $\lab\overline C_{a,a}^\text{noise}\rab$ and $\lab\overline C_{a,b}^\text{noise}\rab$ marked as red X. Fitting $C_1$ and $C_2$ (blue curve) results in overestimation of the true correlation length (green curve) in this case.}
\label{fig:fitting_ex}
\end{figure}

With experimentally obtained fluctuation data one may estimate the normalized auto- and cross-correlation values at the correlation time delay $\tau=0$ denoted as $C_1$ and $C_2$, respectively; while the true normalized correlation values are $\lab\overline C_{a,a}^\text{noise}\rab$ and $\lab\overline C_{a,b}^\text{noise}\rab$ as shown in \reffig{fig:fitting_ex}. Here, $C_1$ is the normalized auto-correlation value after removing the noise by using the technique explained earlier \cite{Zoletnik_PPCF_1998}. As we are restricted to use only two spatial points, one may fit any arbitrary functions to the data. We have chosen to fit a Gaussian function to the experimentally obtained values $C_1$ and $C_2$ motivated by the experimentally observed ion-scale turbulence in tokamaks \cite{Fonck1993, Mckee2003}:
\begin{equation}
\label{eq:gaussian_fit}
C_2 = C_1\exp\lp-\dfrac{\lp x_a-x_b \rp^2}{2\ell_x^2} \rp,
\end{equation}
where we have only one unknown $\ell_x$ which is the \textit{correlation length} that we wish to estimate. The correlation length $\ell_x$ is $\sqrt{2}$ times larger than the characteristic spatial scale $\lx$ (cf. \refeq{eq:corr_anal_simple} and \refeq{eq:norm_corr_noise}). The measured correlation length $\elxM$ is, then, estimated to be
\begin{eqnarray}
\label{eq:two_point_fit}
\elxM = 
\begin{cases}
\infty & C_1 \le C_2  \nonumber \\
\dfrac{\abs{x_a-x_b}}{\sqrt 2} \sqrt{\dfrac{1}{\ln C_1-\ln C_2}} & C_1  > C_2  > 0 \\
0 &  C_2 \le 0.
\end{cases} \\
\end{eqnarray}
The expected value $\mu_{\ell_x}$ and the variance $\sigma^2_{\ell_x}$ of the measured correlation length $\elxM$ are
\begin{eqnarray}
\label{eq:length_mean_var}
\mu_{\ell_x} &=& \int \int \elxM p\lp C_1 \rp p\lp C_2 \rp dC_1 dC_2  \nonumber \\
\sigma^2_{\ell_x} &=& \int \int \lp\elxM\rp^2 p\lp C_1 \rp p\lp C_2 \rp dC_1 dC_2  - \mu^2_{\ell_x}, \nonumber \\
\end{eqnarray}
where $p\lp C_1\rp$ and $p\lp C_2\rp$ are the probability density functions of obtaining $C_1$ and $C_2$, respectively.

\subsection{Correlation length measurement with two spatial points from the analytic expression and synthetic data}
\label{sec:two_point_eq_syn}

\begin{figure}[!]
\includegraphics[width=\linewidth]{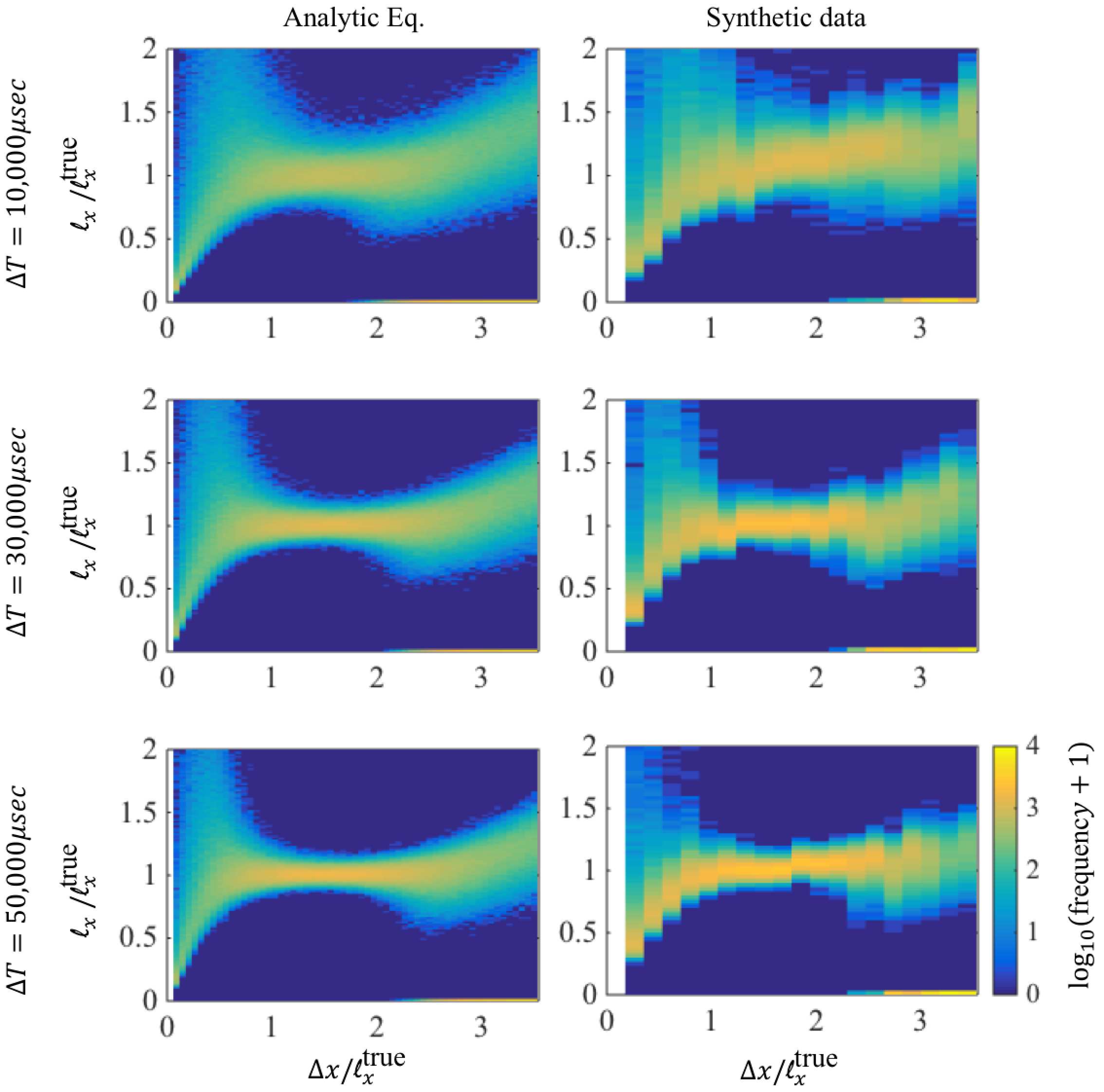}
\caption{Histograms of the normalized correlation length $\elxM/\elxT$ as a function of the separation distance $\Delta x/\elxT$ for different sizes of the total time window $\DT$. Left panels are obtained from the analytic results and right panels from the synthetic data.}
\label{fig:avg_time_hist}
\end{figure}

\begin{figure}[!]
\includegraphics[width=\linewidth]{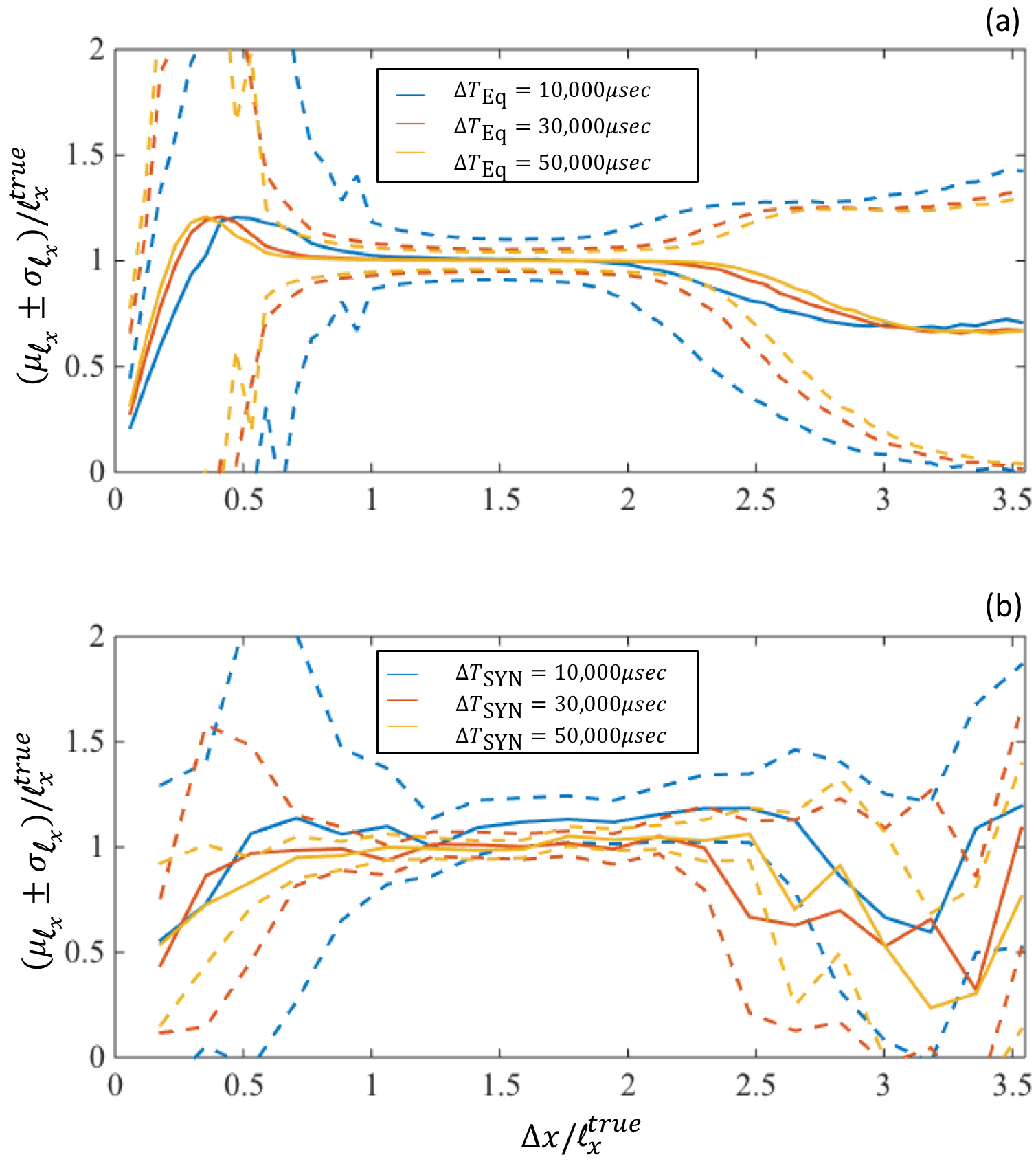}
\caption{Averages (solid lines) and standard deviations from the averages (dashed lines) of the correlation length normalized to $\elxT$ estimated (a) analytically and (b) numerically as a function of the normalized separation distance $\Delta x/\elxT$ for different sizes of the total time window of $\DT= 10,000~\mu$s (blue), $\DT= 30,000~\mu$s (red) and $\DT= 50,000~\mu$s (yellow).}
\label{fig:avg_time_graph}
\end{figure}

To be able to calculate $\mu_{\ell_x}$ and $\sigma^2_{\ell_x}$, we need to know the probability density function of $\lab\NormCorrNoise\rab$, i.e., $p\lp C_1\rp$ and $p\lp C_2\rp$ in \refeq{eq:length_mean_var}. The variance of the ensemble averaged correlation value is $1/M$ times the variance of the correlation value where $M$ is the sample number, i.e., the number of sub-time windows used to measure the ensemble average of the correlation value. Owing to the central limit theorem, this probability density function, $p\lp C_1\rp$ or $p\lp C_2\rp$, is a normal distribution function with the mean of $\lab\NormCorrNoise\rab$ and the standard deviation of $\sigma_{_{\overline C_{a,b}^\text{noise}}}/\sqrt{M}$ which can be estimated either analytically or numerically with the synthetic data.

Once we have the probability density functions $p\lp C_1\rp=P\lp\lab \overline{C}_{a,a}^\text{noise} \rab\rp$ and $p\lp C_2\rp=P\lp \lab \overline{C}_{a,b}^\text{noise} \rab\rp$ either from the analytic results or numerical synthetic data, a Monte-Carlo method by generating $10,000$ samples from these probability density functions is used to generate a histogram of the correlation length normalized to the true correlation length $\elxT=\sqrt{2}\lx$ as a function of the separation distance between the two measured positions $\Delta x$ normalized to the true correlation length $\elxT$ as well.

This histogram is, then, used to estimate the mean $\mu_{\ell_x}$ and the variance $\sigma^2_{\ell_x}$ of the two-point correlation length measurements. Note that we ignore the case of $C_1 \leq C_2$ which corresponds to $\elxM=\infty$. Nevertheless, if $C_1 \leq C_2$ occurs frequently, indicating that the correlation length $\elxM$ is much larger than the separation distance of the two measurement positions, such a signature will be shown as a large variance according to \refeq{eq:two_point_fit} and \refeq{eq:length_mean_var}. Here, we examine two major factors that affect the quality of the correlation length measurements: 1) size of the total time window and 2) noise level.

To quantify how the size of the total time window affects the measured correlation length, we generate three sets of synthetic data with $\lx = 0.3$ m, $\tlife=15~\mu$s, $v=5,000$ m/s and $\DTsub=585~\mu$s at the fixed FNR of $0.5$, whereas $\DT$ are set to be $10,000~\mu$s, $30,000~\mu$s and $50,000~\mu$s. \reffig{fig:avg_time_hist} shows the histograms of the obtained correlation length with the $p\lp C_1\rp$ and $p\lp C_2\rp$ from the analytic results (left panels) and the synthetic data (right panels). Based on these histograms, we finally obtain the average and standard deviation of the correlation length (normalized to $\elxT$) as a function of $\Delta x/\elxT$ as shown in \reffig{fig:avg_time_graph}. Although \reffig{fig:avg_time_graph} is more succinct than \reffig{fig:avg_time_hist}, it is better to consider \reffig{fig:avg_time_hist} because the first ($\mu_{\ell_x}$) and second ($\sigma^2_{\ell_x}$) moments become less representative as the histogram deviates from a normal distribution, especially for small and large $\Delta x/\elxT$.

The similar trend between the analytic and numerical results shows that increasing the size of the total window decreases the standard deviation of the measurements, and the two-point correlation length measurement becomes less reliable if the separation distance between the two measured points are either smaller than or more than two times larger than the actual correlation length of the fluctuation data.

\begin{figure}[!]
\includegraphics[width=\linewidth]{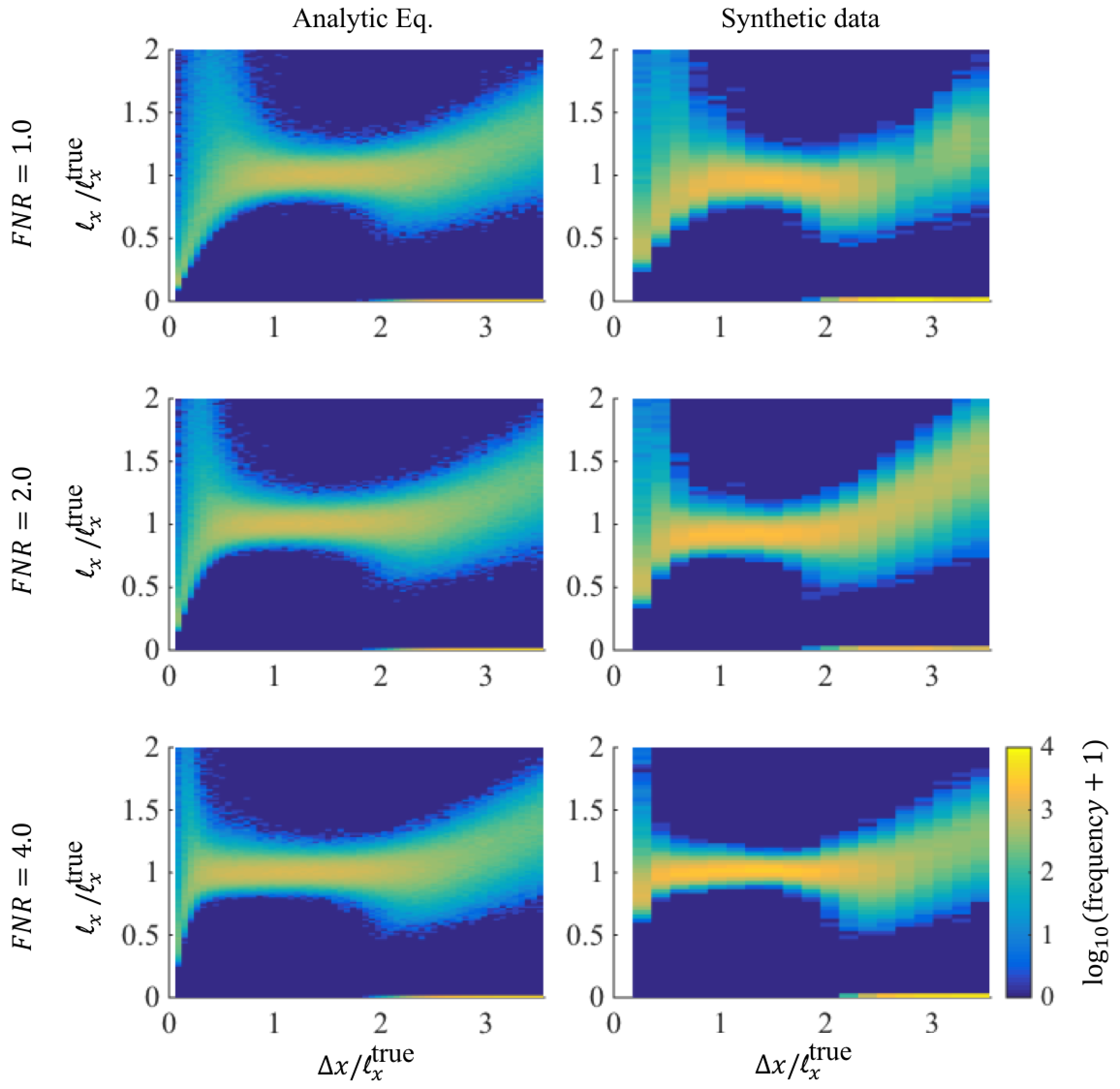}
\caption{Same as \reffig{fig:avg_time_hist} for a fixed $\DT = 15,000~\mu$s with the different values of FNR: 1.0 (top), 2.0 (middle) and 4.0 (bottom).}
\label{fig:noise_level_hist}
\end{figure}

\begin{figure}[!]
\includegraphics[width=\linewidth]{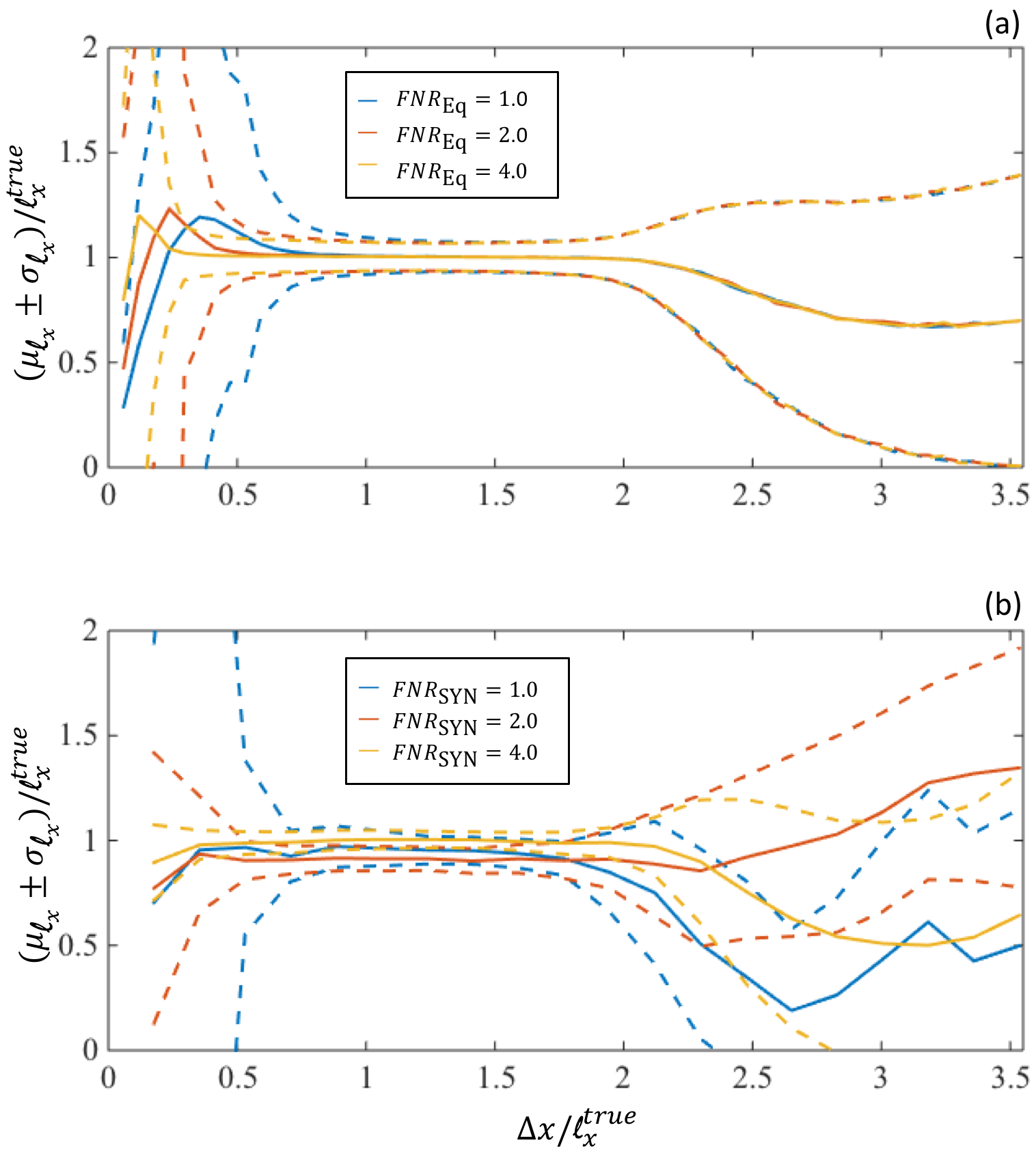}
\caption{Same as \reffig{fig:avg_time_graph} for a fixed $\DT = 15,000~\mu$s with the different values of FNR: 1.0 (blue), 2.0 (red) and 4.0 (yellow).}
\label{fig:noise_level_graph}
\end{figure}

Next, we examine the effect of noise level, i.e., the fluctuation-to-noise ratio (FNR), on the two-point correlation length measurements. Three sets of synthetic data with $\lx = 0.3$ m, $\tlife=15~\mu$s, $v=5,000$ m/s, $\DT = 15,000~\mu$s and $\DTsub=585~\mu$s are generated with different values of FNR: $1.0$, $2.0$ and $4.0$. \reffig{fig:noise_level_hist} shows the histograms of the obtained correlation length using the analytic results (left panel) and synthetic data (right panel), whereas the average and standard deviation of the correlation length is shown in \reffig{fig:noise_level_graph}.

We find that the lower limit on the $\Delta x/\elxT$ is dependent on the FNR; while the upper limit is not. This observation can be explained based on \reffig{fig:correlation_comparison}. We see that the standard deviation levels are larger at small $\Delta x$ for the smaller FNR. This results in a larger value on the lower limit of $\Delta x/\elxT$ with the smaller FNR. In addition, larger standard deviation at the small $\Delta x/\elxT$ is more probable to have $C_1\le C_2$ in \refeq{eq:two_point_fit}. \reffig{fig:correlation_comparison} also shows that although the standard deviation levels are larger with the larger FNR at large $\Delta x$, they are smaller at $\Delta x=0$. Therefore, we observe similar upper limit on $\Delta x/\elxT$ for the investigated FNR cases.

\subsection{Unnormalized vs. normalized correlation function for the correlation length measurement}
\label{sec:unnormalized_vs_normalized}

\begin{figure}[!]
\includegraphics[width=\linewidth]{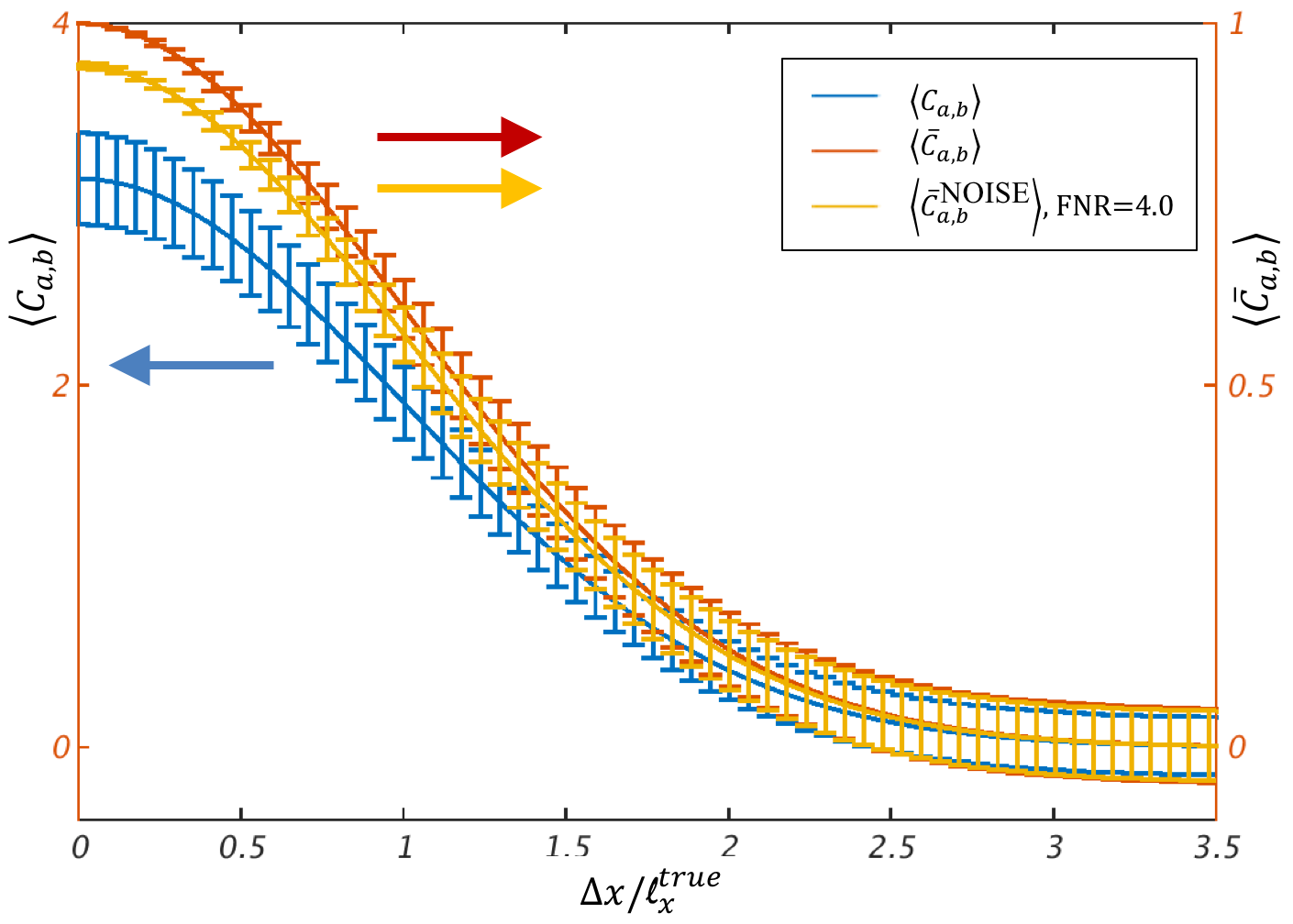}
\caption{Unnormalized (blue) and normalized correlation functions with no noise (red) and noise level of FNR=4.0 (yellow) as a function of $\Delta x/\elxT$ calculated with the analytic results. As the $\Delta x/\elxT$ approaches to zero, error bars become smaller for the normalized correlation functions than the unnormalized ones. Here, the error bars represent the standard deviation of the mean, i.e., $1/\sqrt{M}$ times the standard deviation of the correlation value.}
\label{fig:cov_vs_corr}
\end{figure}

The variance of the correlation length $\sigma^2_{\ell_x} $ is a function of $p\lp C_1 \rp$ and $p\lp C_2 \rp$ as shown in \refeq{eq:length_mean_var}, i.e., the narrower the $p\lp C_1 \rp$ and $p\lp C_2 \rp$, the smaller the variance. Considering an extreme case of absolutely no noise, we see that $p\lp C_1 \rp$ is a delta function, i.e., no variance, for the normalized correlation function (see \refeq{eq:var_of_norm_corr}) while it always has a non-zero width, i.e., non-zero variance, for the unnormalized correlation function (see \refeq{eq:var_of_corr_approx}). \reffig{fig:cov_vs_corr} shows the unnormalized (blue) and normalized correlation functions with no noise (red) and FNR=4.0 (yellow) as a function of $\Delta x/\elxT$ calculated with the analytic results. Here, the error bars represent the standard deviation of the mean, i.e., $1/\sqrt{M}$ times the standard deviation of the correlation value; thus, they provide approximate widths of $p\lp C_1 \rp$ and $p\lp C_2 \rp$. Direct comparison of the widths of $p\lp C_2 \rp$ between unnormalized and normalized correlation functions can be tricky as they are functions of $\Delta x$, but the very small width of the $p\lp C_1 \rp$ at $\Delta x=0$ for the normalized correlation function, in general, provides better correlation length measurements compared to the unnormalized correlation function.

\begin{figure}[!]
\includegraphics[width=\linewidth]{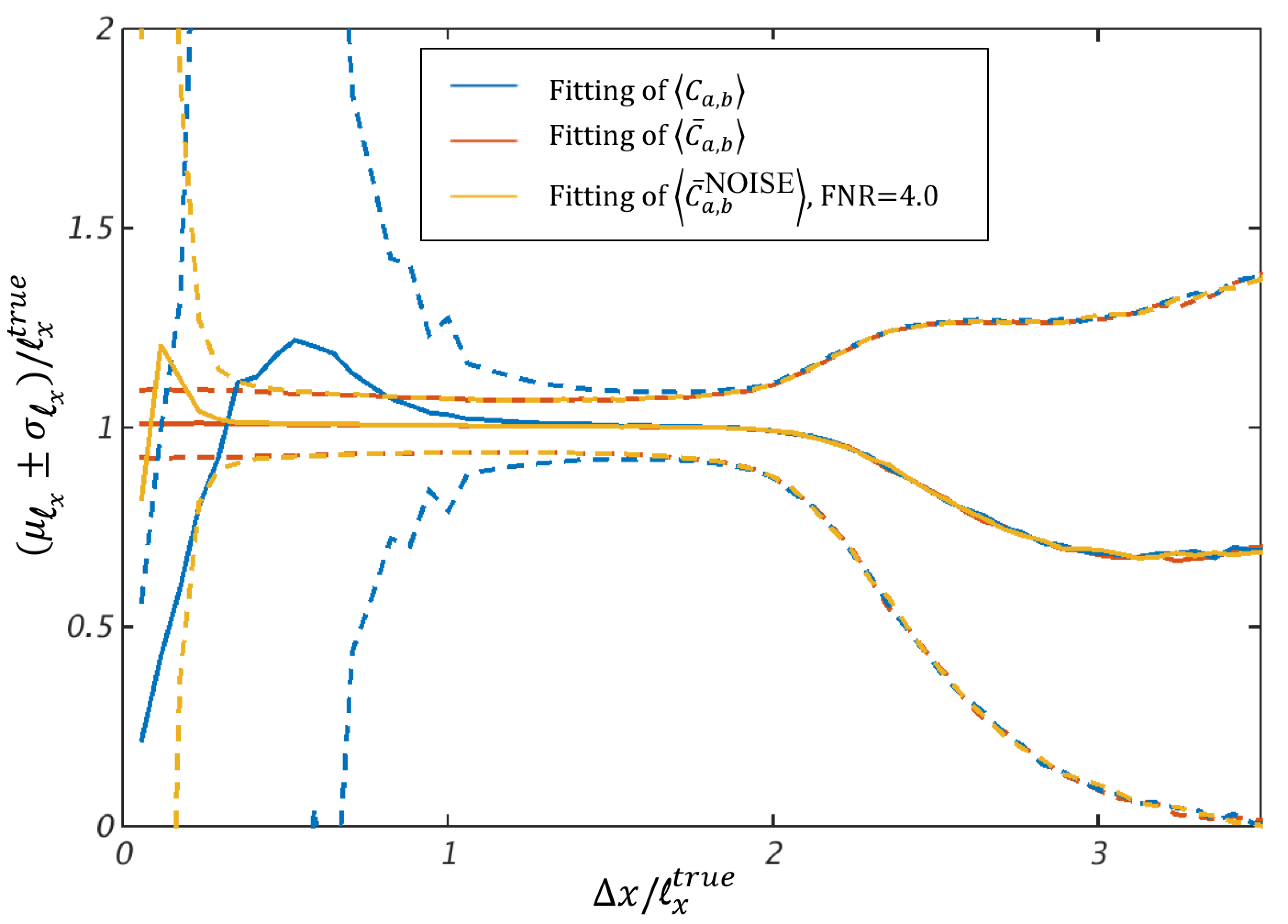}
\caption{The two point correlation length measurement using normalized correlation function (blue line) and unnormalized correlation function (red line) when the noise level is zero. Solid lines show the mean and dashed lines show the standard deviation.}
\label{fig:cov_vs_corr_fit}
\end{figure}

\reffig{fig:cov_vs_corr_fit} shows the average and standard deviation of the correlation length estimated with the unnormalized (blue) and normalized correlation functions without noise (red) and with the noise level of FNR=4.0 (yellow) with the analytic results as in \reffig{fig:cov_vs_corr}. We see that the normalized correlation function provides smaller standard deviation and bias error of the measured correlation length especially for $\Delta x \le \elxT$.

\subsection{Experimental data: two point correlation length measurement for real density fluctuations}
\label{sec:real_case}

Reliability and accuracy of the two-point correlation length measurement have been examined analytically and numerically. We now apply the two-point correlation length measurement on the density fluctuation data experimentally obtained via the BES system \cite{Lampert2015} installed in KSTAR. 

\begin{figure}[!]
\includegraphics[width=\linewidth]{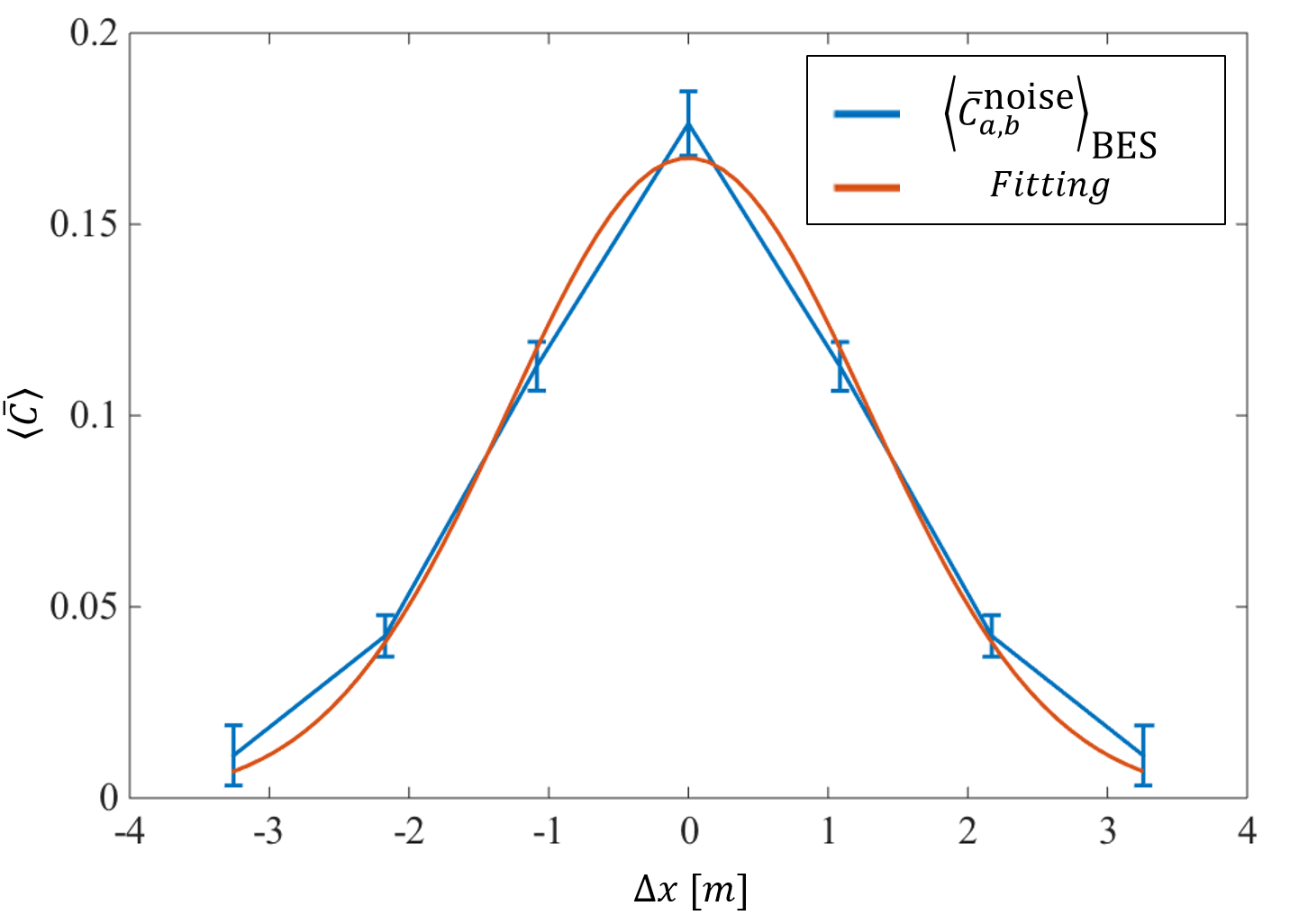}
\caption{An example of a Gaussian fitting (red) to the correlation values (blue) from poloidally aligned 4 detectors for KSTAR shot \#9133. Correlation values for $\Delta x<0$ are copies of the values from $\Delta x>0$ to aid visualization. This Gaussian fitting provides the `true' poloidal correlation length $\ell_{\theta\text{(4 points)}}$.}
\label{fig:4points}
\end{figure}

Since the `true' correlation length of the fluctuation data are not available, we estimate the correlation length using the four points (poloidally aligned four detectors) denoted as $\ell_{\theta\text{(4 points)}}$ and treat this value as the `true' correlation length. \reffig{fig:4points} shows an example of four-point measurement of the correlation values (blue) at the time delay $\tau=0$. Note that correlation values for $\Delta x<0$ are just copies of the values from $\Delta x>0$ to aid visualization. Gaussian fitting (red) to the correlation values provides the correlation length $\ell_{\theta\text{(4 points)}}$.

\begin{figure}[!]
\includegraphics[width=\linewidth]{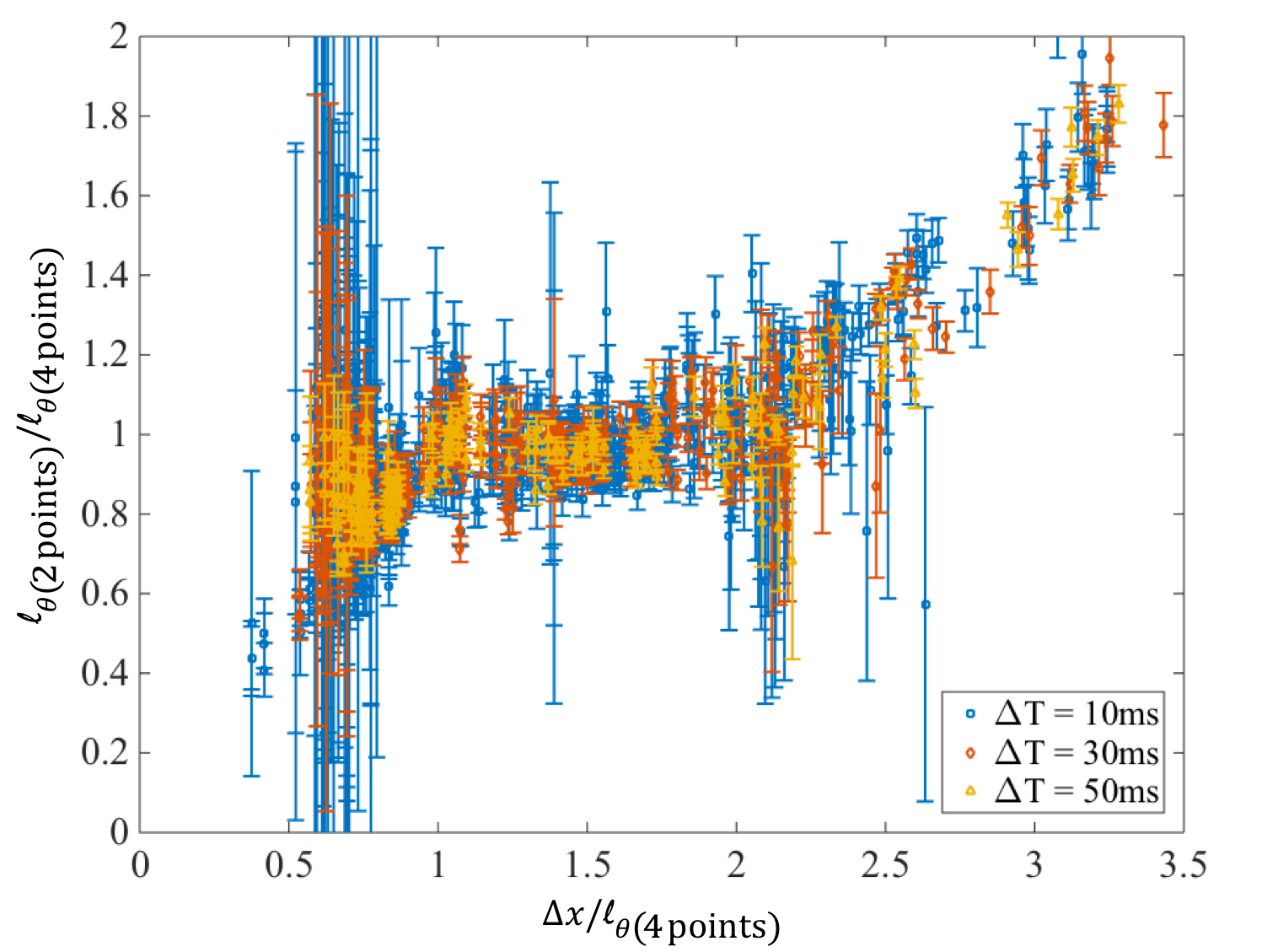}
\caption{Two-point correlation length measurement $\ell_{\theta\text{(2 points)}}$ normalized to four-point correlation length measurement $\ell_{\theta\text{(4 points)}}$ as a function of the separation distance $\Delta x/\ell_{\theta\text{(4 points)}}$ from the KSTAR BES data for shot \#9133. The result is quantitatively similar to \reffig{fig:avg_time_hist} (or \reffig{fig:avg_time_graph}).}
\label{fig:two_point_measure_real}
\end{figure}

Then, we estimate the correlation length using only two measurement points denoted as $\ell_{\theta\text{(2 points)}}$. \reffig{fig:two_point_measure_real} shows $\ell_{\theta\text{(2 points)}}/\ell_{\theta\text{(4 points)}}$ as a function of $\Delta x/\ell_{\theta\text{(4 points)}}$ for different sizes of the total time window $\DT=10$ ms (blue), $30$ ms (red) and $50$ ms (yellow). The result is quantitatively similar to \reffig{fig:avg_time_hist} (or \reffig{fig:avg_time_graph}), i.e., results obtained analytically and numerically: two-point correlation length measurements are not biased for $1\le \Delta x/\ell_{\theta\text{(4 points)}}\le2$, and a larger size of total time window provides smaller variance. 

\section{Conclusion}
\label{sec:conclusion}

We have analytically derived the normalized and unnormalized correlation functions of Gaussian shaped moving fluctuation data as well as their associated variances. The analytic results are found to have good agreement with the numerical results. Based on the correlation functions, the reliability and accuracy of the two-point correlation length measurement have been examined, where we have found that the separation distance between the two measurements points needs to be within one and two times of the correlation length to obtain reliable and accurate true results. The lower limit is found to be dependent on the ratio of fluctuation to noise. This indicates that if an obtained correlation length from two-point measurements are much larger or smaller than the separation distance, validity of the obtained correlation length must be questioned. Our results can also be used as design criteria when one builds a diagnostic system measuring fluctuations of any physical quantities with an aim of obtaining correlation lengths.

\begin{acknowledgments}
This work is supported by National R\&D Program through the National Research Foundation of Korea (NRF) funded by the Ministry of Science, ICT \& Future Planning (grant number 2014M1A7A1A01029835) and the KUSTAR-KAIST Institute, KAIST, Korea.
\end{acknowledgments}

\appendix

\section{Variance of the correlation value}
\label{app:var_corr_value}

The variance of the estimated correlation value $\varCorr$ by definition is
\begin{equation}
\label{eq:var_of_corr_definition}
\varCorr=\lab C_{a,b}^2 \rab - \lab C_{a,b} \rab^2.
\end{equation}
By invoking a similar approach done by Kim et al. \cite{jwkim_cpc_2016}, we find
\begin{widetext}
\begin{eqnarray}
\label{eq:var_of_corr}
\varCorr &\approx& -N A^4 \pi^2 \dfrac{\lx^2}{\DL^2} \dfrac{\tlife^2}{\DT^2} \exp\lsb -\dfrac{(x_a-x_b)^2}{2\lx^2} \rsb  \nonumber\\ 
&& - 4 N(N-1)A^4 \pi^3 \dfrac{\lx^2}{\DL^2} \dfrac{\tlife^2}{\DT^2} \dfrac{\tac^2}{\DTsub^2}  \nonumber\\
&& -4N^2 A^4 \pi^3 \dfrac{\lx^2}{\DL^2} \dfrac{\tlife^2}{\DT^2} \dfrac{\tac^2}{\DTsub^2} \exp\lsb -\dfrac{(x_a-x_b)^2}{2(\lx^2+\tlife^2v^2)} \rsb \nonumber\\ 
&& +\sqrt{2} N(N-1) A^4 \pi^{5/2} \dfrac{\lx^2}{\DL^2} \dfrac{\tlife^2}{\DT^2} \dfrac{\tac}{\DTsub} \left( 1+ \exp\lsb -\dfrac{(x_a-x_b)^2}{2\lx^2} \rsb \right) \nonumber\\ 
&& + 4 N A^4 \pi^{5/2} \dfrac{\lx^2}{\DL^2} \dfrac{\tlife^2}{\DT^2} \dfrac{\tac}{\DTsub} \exp\lsb -\dfrac{(2\lx^2+\tlife^2v^2)(x_a-x_b)^2}{4\lx^2(\lx^2+\tlife^2v^2)} \rsb \nonumber\\ 
&& + 6\sqrt{2} N A^4 \pi^{5/2} \dfrac{\lx}{\DL} \dfrac{\tlife}{\DT} \dfrac{\tac^3}{\DTsub^3} \exp\lsb -\dfrac{(x_a-x_b)^2}{4(\lx^2+\tlife^2v^2)} \rsb \nonumber\\ 
&& - 6\sqrt{2} N A^4 \pi^{2} \dfrac{\lx}{\DL} \dfrac{\tlife}{\DT} \dfrac{\tac^2}{\DTsub^2} \exp\lsb -\dfrac{(2\lx^2+\tlife^2v^2)(x_a-x_b)^2}{4\lx^2(\lx^2+\tlife^2v^2)} \rsb \nonumber\\ 
&& + \dfrac{3}{\sqrt{2}} N A^4 \pi^{3/2} \dfrac{\lx}{\DL} \dfrac{\tlife}{\DT} \dfrac{\tac}{\DTsub} \exp\lsb -\dfrac{(x_a-x_b)^2}{2\lx^2} \rsb,
\end{eqnarray}
\end{widetext}
where the approximation is for $\DL\gg\lx$, $\DT\gg\DTsub\gg\tlife$ (or $\tac$) consistent with the conditions on obtaining $\lab C_{a,b}\rab$ in \refeq{eq:corr_anal_final}. Then, having the spatio-temporal filling factor $F\sim\mathcal{O}\lp 1\rp$ and assuming $N\gg 1$ which is well satisfied with large $\DL$ and $\DT$, all the terms are negligible compared to the fourth and the eighth terms on the right-hand-side of \refeq{eq:var_of_corr}. Note that $N\lp N-1\rp$ is approximated to $N^2$ in the fourth term to get \refeq{eq:var_of_corr_approx}.

\section{Ensemble average approximation of the normalized correlation function}
\label{app:ensemble_average_approx}
Here, we explain the rationale and the assumptions we have made in $\lab \frac{C_{a,b}}{\delta_f^2} \rab \approx \frac{\lab C_{a,b}\rab}{\lab \delta_f^2 \rab}$ which is the approximation in the second line of the right-hand-side of \refeq{eq:corr_anal_simple}.

Suppose there are two random variables $x$ and $y$ as the realizations of $X$ and $Y$, respectively. For any $f\lp x, y\rp$, the bivariate first order Taylor expansion about $\lp x_1, y_1\rp$ is 
\begin{eqnarray}
\label{eq:bivariate_taylor}
f(x,y)&=&f(x_1,y_1)\nonumber \\ \nonumber
&& + \left. \dfrac{\partial f(x,y)}{\partial x}\right|_{(x_1,y_1)} (x-x_1) \\ \nonumber
&& + \left. \dfrac{\partial f(x,y)}{\partial y}\right|_{(x_1,y_1)} (y-y_1)\\
&& + \text{ remainder}.
\end{eqnarray}
Applying \refeq{eq:bivariate_taylor} with $f\lp x, y\rp=x/y$ for $x=C_{a,b}$ and $y= \delta_f^2$, the ensemble average of $f\lp x, y\rp$ about $x_1=\lab X\rab$ and $y_1=\lab Y\rab$ is
\begin{eqnarray}
\label{eq:expect_bivariate_taylor}
\lab \frac{C_{a,b}}{\delta_f^2} \rab & = & \lab f\lp x,y\rp\rab \nonumber \\
& = &\lab f\lp\lab X\rab, \lab Y\rab\rp \rab \nonumber \\ 
&& +  \dfrac{\partial f\lp x,y \rp}{\partial x}\bigg|_{\lp\lab X\rab, \lab Y\rab\rp} \underbrace{\lab x-\lab X\rab \rab}_{=0} \nonumber \\ 
&& +  \dfrac{\partial f\lp x,y \rp}{\partial y}\bigg|_{\lp\lab X\rab, \lab Y\rab\rp} \underbrace{\lab y-\lab Y\rab\rab}_{=0} \nonumber \\
&& + \text{ remainder} \nonumber \\
&\approx& \lab \frac{\lab C_{a,b}\rab}{\lab\delta_f^2\rab} \rab = \frac{\lab C_{a,b}\rab}{\lab\delta_f^2\rab}.
\end{eqnarray}
Ignoring the remainder, i.e., the higher order terms, in the last line is valid when the standard deviation of a random variable ($\sigma_y$) is small compared to its mean (or expectation) value ($\lab Y\rab$) \cite{Johnson}. Note that the higher order terms in $x$, i.e., $\partial^n/\partial x^n$ for $n\ge 2$, are zeros for $f\lp x, y\rp = x/y$. Thus, the ensemble average approximation $\lab \frac{C_{a,b}}{\delta_f^2} \rab \approx \frac{\lab C_{a,b}\rab}{\lab \delta_f^2 \rab}$ is valid for $\sigma_{C_{a,a}} = \sigma_{\delta^2_f}  \ll \lab \delta^2_f\rab = \lab C_{a,a}\rab$ which is satisfied for $\tlife\ll\DTsub$ according to \refeq{eq:corr_anal_final} and \refeq{eq:var_of_corr_approx}. 

\section{Variance of the normalized correlation value with and without noise}
\subsection{Without noise}
\label{app:var_norm_corr_value}

To derive the variance of the normalized correlation value $\varNormCorr$, we invoke an approximated form of the variance of products \cite{goodman_1960} for $g=A/B$, namely
\begin{equation}
\label{eq:uncertainty_propagation_quotient}
\sigma_g^2 \approx g^2 \lsb \lp\dfrac{\sigma_A}{A}\rp^2 + \lp\dfrac{\sigma_B}{B}\rp^2 - 2\rho_{AB}\dfrac{\sigma_A\sigma_B}{AB}\rsb,
\end{equation}
where $\sigma_g$, $\sigma_A$ and $\sigma_B$ are the standard deviations of $g$, $A$ and $B$, respectively. $\rho_{AB}$ is the normalized correlation between $A$ and $B$. For our case, $g=\lab\overline C_{a,b}\rab$, $A=\lab C_{a,b}\rab$ and $B=\lab\delta_a \delta_b\rab=\lab\delta_f^2\rab$ with $\sigma^2_A = \varCorr$ and $\sigma^2_B = \sigma^2_{\delta_f^2}$. A difficulty to derive $\varNormCorr$ arises due to the the correlation between $C_{a,b}$ and $\delta_f^2$, i.e., $\rho_{AB}$. For this reason, we take two extreme cases where $\rho_{AB}\approx 1$ and $\rho_{AB}\approx 0$, and connect them smoothly based on an educated guess. Note that we confirm numerically in \refsec{sec:compare_eq_syn} that our result is indeed valid.

For $x_a=x_b$, we know that $C_{a,b}=C_{a,a}=\delta_f^2$, i.e., $A=B$, resulting in $\rho_{AB}=1$. This simply gives $\sigma^2_g=0$, i.e., $\varNormCorr=0$. This result is what we expect since we know that the normalized auto-correlation at the time delay $\tau=0$ is unity by definition, and there is no variance associated with it if the signal contains no noise.

The opposite extreme case is obtained by considering two spatial positions where $\abs{x_a-x_b}\gg\lx$ such that there is absolutely no correlation between the signals at $x_a$ and $x_b$, i.e., $C_{a,b}\approx 0$. Thus, there is no correlation between $C_{a,b}$ and $\delta_f^2$, resulting in $\rho_{AB}\approx 0$. For this case, $\varNormCorr$ becomes
\begin{eqnarray}
\varNormCorr &\approx& \lp\dfrac{\lab C_{a,b}\rab}{\lab\delta_f^2\rab}\rp^2 \lsb \lp\dfrac{\sigma_{_{C_{a,b}}}}{\lab C_{a,b}\rab}\rp^2 + \lp\dfrac{\sigma_{\lab\delta_f\rab^2} }{\lab\delta_f^2\rab}\rp^2\rsb \nonumber \\
&\approx& \lp\dfrac{\sigma_{_{C_{a,b}}}}{\lab\delta_f^2\rab} \rp^2 = \dfrac{\varCorr}{\pi^2 A^4 F^2}, 
\end{eqnarray}
where we have ignored the second term inside the square bracket for very small $\lab C_{a,b}\rab$ to get the last line and used the fact that $\lab\delta_f^2\rab=\pi A^2 F$.

With these two results on $\varNormCorr$ for $\abs{x_a-x_b}=0$ and $\abs{x_a-x_b}\gg\lx$, we connect them with a Gaussian shape motivated by the form of variance for the unnormalized correlation value $\varCorr$ as in \refeq{eq:var_of_corr_approx}. Thus, we finally obtain the variance of the normalized correlation value as:
\begin{equation}
\varNormCorr\approx \dfrac{\varCorr}{\pi^2 A^4 F^2}\lp 1-\exp\lsb -\dfrac{(x_a-x_b)^2}{2\lx^2}  \rsb \rp^2,
\end{equation}
which is what we have in \refeq{eq:var_of_norm_corr}. 

\subsection{With noise}
\label{app:var_norm_corr_value_noise}

To obtain the variance of the normalized correlation value with noise $\varNormCorrNoise$, we again use an approximated form of the variance of products \cite{goodman_1960} for $g=AB$ which is
\begin{equation}
\label{eq:uncertainty_propagation_product}
\sigma_g^2 \approx g^2 \lsb \lp\dfrac{\sigma_A}{A}\rp^2 + \lp\dfrac{\sigma_B}{B}\rp^2 + 2\rho_{AB}\dfrac{\sigma_A\sigma_B}{AB}\rsb,
\end{equation}
where $g=\lab \NormCorrNoise\rab$, $A=\lab\delta_f^2\rab/\lab\delta_f^2+\delta_n^2\rab$ and $B=\lab\overline C_{a,b} \rab$ with $\sigma_g^2=\varNormCorrNoise$, $\sigma_A^2=\sigma^2_{\delta_f^2/\lp\delta_f^2+\delta_n^2\rp}$ and $\sigma_B^2 = \varNormCorr$. $\rho_{AB}$ is the normalized correlation between $\delta_f^2/\lp\delta_f^2+\delta_n^2\rp$ and $\overline C_{a,b}$ which is zero because the normalized correlation value $\overline C_{a,b}$ is independent of the fluctuation level.

Since $\sigma^2_{\delta_f^2/\lp\delta_f^2+\delta_n^2\rp}$ is unknown, we calculate it using \refeq{eq:uncertainty_propagation_quotient} where the variances of $\delta_f^2$ and $\delta_n^2$ are $\sigma^2_{\delta_f^2}$ and $\sigma^2_{\delta_n^2}$, respectively. Then, the variance of $\delta_f^2 + \delta_n^2$ denoted as $\sigma^2_{\delta_f^2 + \delta_n^2}$ is $\sigma^2_{\delta_f^2}+\sigma^2_{\delta_n^2}$ because $\delta_f^2$ and $\delta_n^2$ are uncorrelated. Furthermore, the normalized correlation between $\lab \delta_f^2\rab$ and $\lab \delta_f^2+\delta_n^2\rab$ are approximately unity since $\delta_n^2$ is a constant as long as the sub-time window size $\DTsub$ is much larger than the auto-correlation time of the noise which is a sampling time. Thus, $\sigma^2_{\delta_f^2/\lp\delta_f^2+\delta_n^2\rp}$ is
\begin{eqnarray}
\label{eq:var_of_auto_with_noise}
\sigma^2_{\delta_f^2/\lp\delta_f^2+\delta_n^2\rp}&\approx&\lp\dfrac{\lab\delta_f^2\rab}{\lab\delta_f^2+\delta_n^2\rab}\rp^2 \nonumber \\
&& \lsb\dfrac{\sigma^2_{\delta_f^2}}{\lab\delta_f^2\rab^2}  + \dfrac{\sigma^2_{\delta_f^2+\delta_n^2}}{\lab\delta_f^2+\delta_n^2\rab^2} -2\dfrac{\sigma_{\delta_f^2}\sigma_{\delta_f^2+\delta_n^2} }{\lab\delta_f^2\rab \lab\delta_f^2+\delta_n^2\rab} \rsb \nonumber \\
&=&\lp\dfrac{\lab\delta_f^2\rab}{\lab\delta_f^2\rab+\lab\delta_n^2\rab} \rp^2 \lsb \dfrac{\sigma_{\delta_f^2}}{\lab\delta_f^2\rab}-\dfrac{\sqrt{\sigma^2_{\delta_f^2}+\sigma^2_{\delta_n^2} }}{ \lab\delta_f^2\rab+\lab\delta_n^2\rab }  \rsb^2.  \nonumber \\
\end{eqnarray}
We, then, obtain $\varNormCorrNoise$ as
\begin{eqnarray}
\label{eq:var_of_norm_corr_noise_app}
\varNormCorrNoise&\approx&\lp  \dfrac{\lab\delta_f^2\rab}{\lab\delta_f^2\rab+\lab\delta_n^2\rab} \lab\overline C_{a,b} \rab  \rp^2\times \nonumber \\
&&\lsb \lp\dfrac{\sigma_{\delta_f^2}}{\lab\delta_f^2\rab}-\dfrac{\sqrt{\sigma^2_{\delta_f^2}+\sigma^2_{\delta_n^2} }}{ \lab\delta_f^2\rab+\lab\delta_n^2\rab }\rp^2
+\dfrac{ \varNormCorr }{ \lab\overline C_{a,b} \rab^2  }\rsb \nonumber \\
&\approx&\lab\NormCorrNoise\rab^2 \nonumber \\
&&\lsb \lp\dfrac{\sigma_{\delta_f^2}}{\lab\delta_f^2\rab}-\dfrac{\sigma_{\delta_f^2}}{ \lab\delta_f^2\rab+\lab\delta_n^2\rab }\rp^2+\dfrac{ \varNormCorr }{ \lab\overline C_{a,b} \rab^2  }\rsb, \nonumber \\
\end{eqnarray}
where we assume that $\sigma^2_{\delta_f^2}\gg\sigma^2_{\delta_n^2}$ in the last line approximation since $\delta_n^2$ is almost a constant. This is the \refeq{eq:var_of_norm_corr_noise}. \reffig{fig:correlation_comparison} shows how our analytical result agrees with the numerically estimated variances using the synthetic data including the noise effects.

\section{The two-point measurement of the correlation length for the Lorentzian-shaped eddies}
\label{app:Lorentzian}

As the Lorentzian-shaped eddies at the edge (SOL) of the magnetically confined plasmas are observed \cite{Maggs_PRL_2011, Hornung_PoP_2011, DIppolito_PoP_2011}, we discuss accuracy and reliability of the two-point measurement of the correlation length for the Lorentzian-shaped eddies in this section. Fluctuating signals are the sum of eddies as in \refeq{eq:eddy_sum} with moving Lorentzian eddies:
\begin{eqnarray}
\label{eq:cauchy_eddy}
S_{a_i}\lp t\rp = A_i && \dfrac{1}{\lp \dfrac{t-t_i}{\tlife} \rp^2+1} \nonumber \\
&&\times \dfrac{1}{\lp \dfrac{x_a-v(t-t_i)^2-x_i}{\lx} \rp^2+1},
\end{eqnarray}
where $A_i$ is selected from a normal distribution; while $x_i$ and $t_i$ are generated from uniform distributions. 

\begin{figure}[t]
\includegraphics[width=\linewidth]{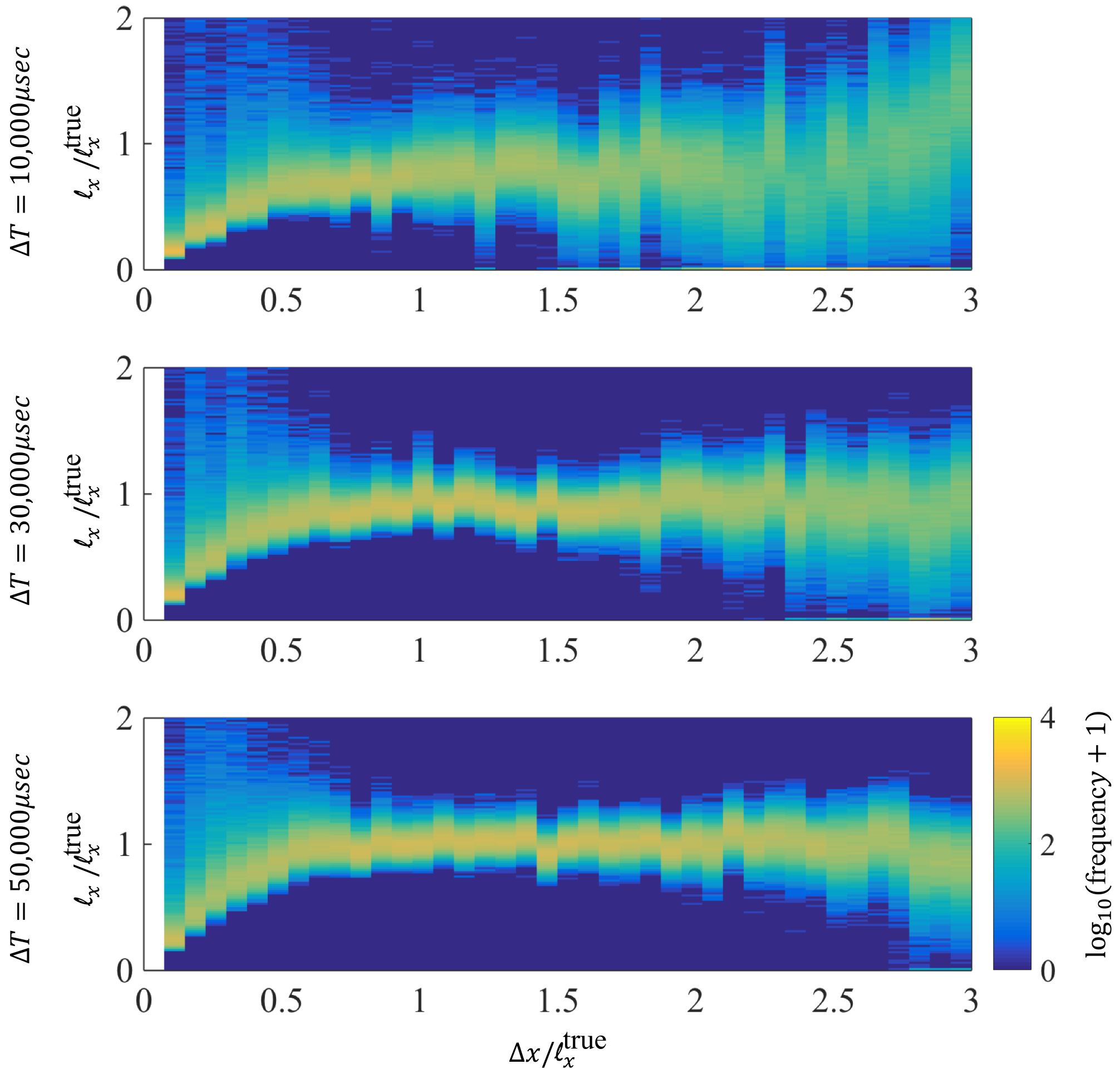}
\caption{Histograms of the normalized correlation length $\elxM/\elxT$ as a function of the separation distance $\Delta x/\elxT$ for different sizes of the total time window $\DT$. They are obtained from the synthetic data.}
\label{fig:avg_time_hist_cauchy}
\end{figure}

\begin{figure}[t]
\includegraphics[width=\linewidth]{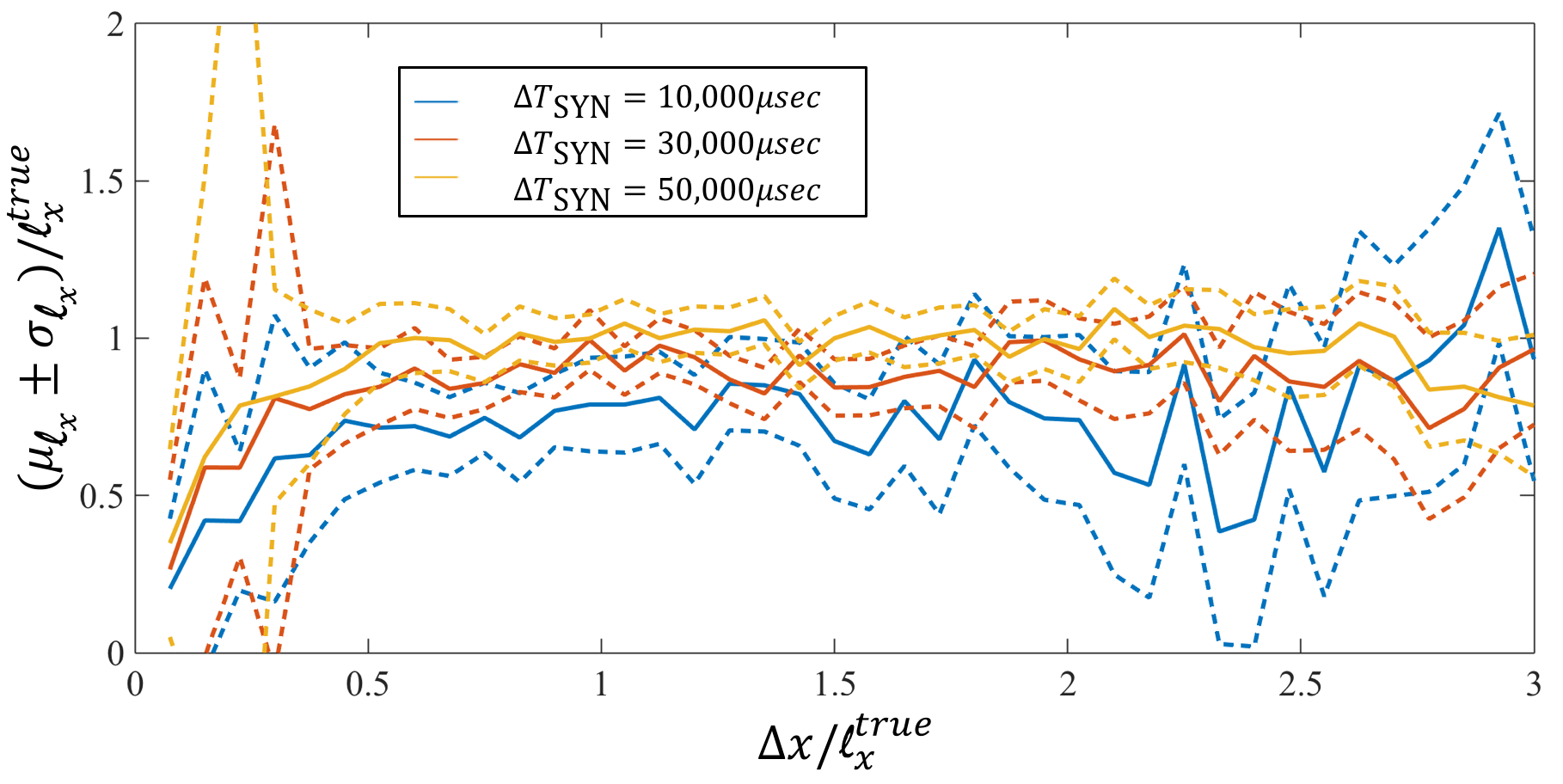}
\caption{Averages (solid lines) and standard deviations (dashed lines) of the correlation length normalized to $\elxT$ estimated numerically as a function of the normalized separation distance $\Delta x/\elxT$ for different sizes of the total time window of $\DT= 10,000~\mu$s (blue), $\DT= 30,000~\mu$s (red) and $\DT= 50,000~\mu$s (yellow).}
\label{fig:avg_time_graph_cauchy}
\end{figure}

\begin{figure}[t]
\includegraphics[width=\linewidth]{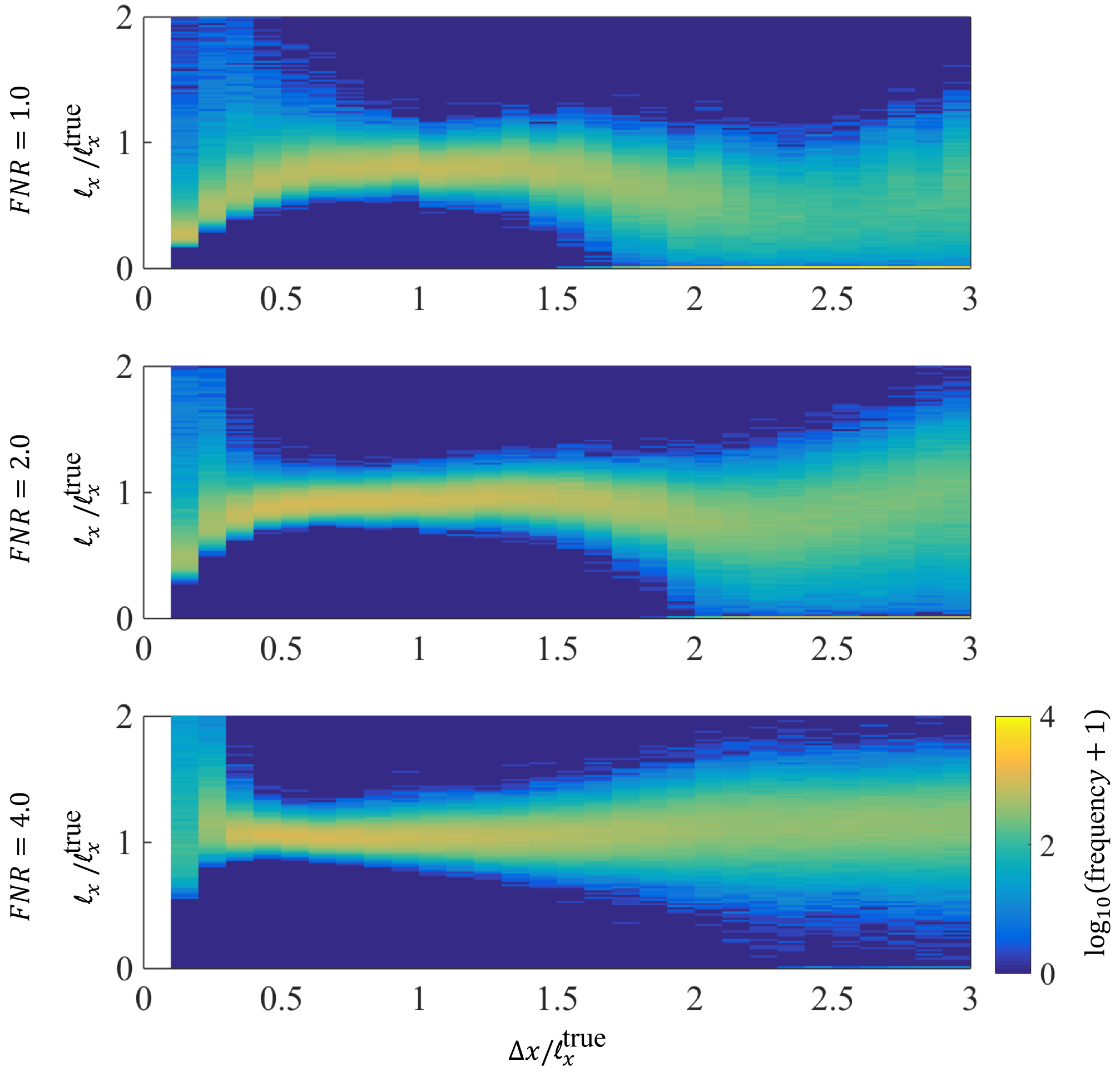}
\caption{Same as \reffig{fig:avg_time_hist_cauchy} for a fixed $\DT = 15,000~\mu$s with the different values of FNR: 1.0 (top), 2.0 (middle) and 4.0 (bottom).}
\label{fig:noise_level_hist_cauchy}
\end{figure} 

\begin{figure}[t]
\includegraphics[width=\linewidth]{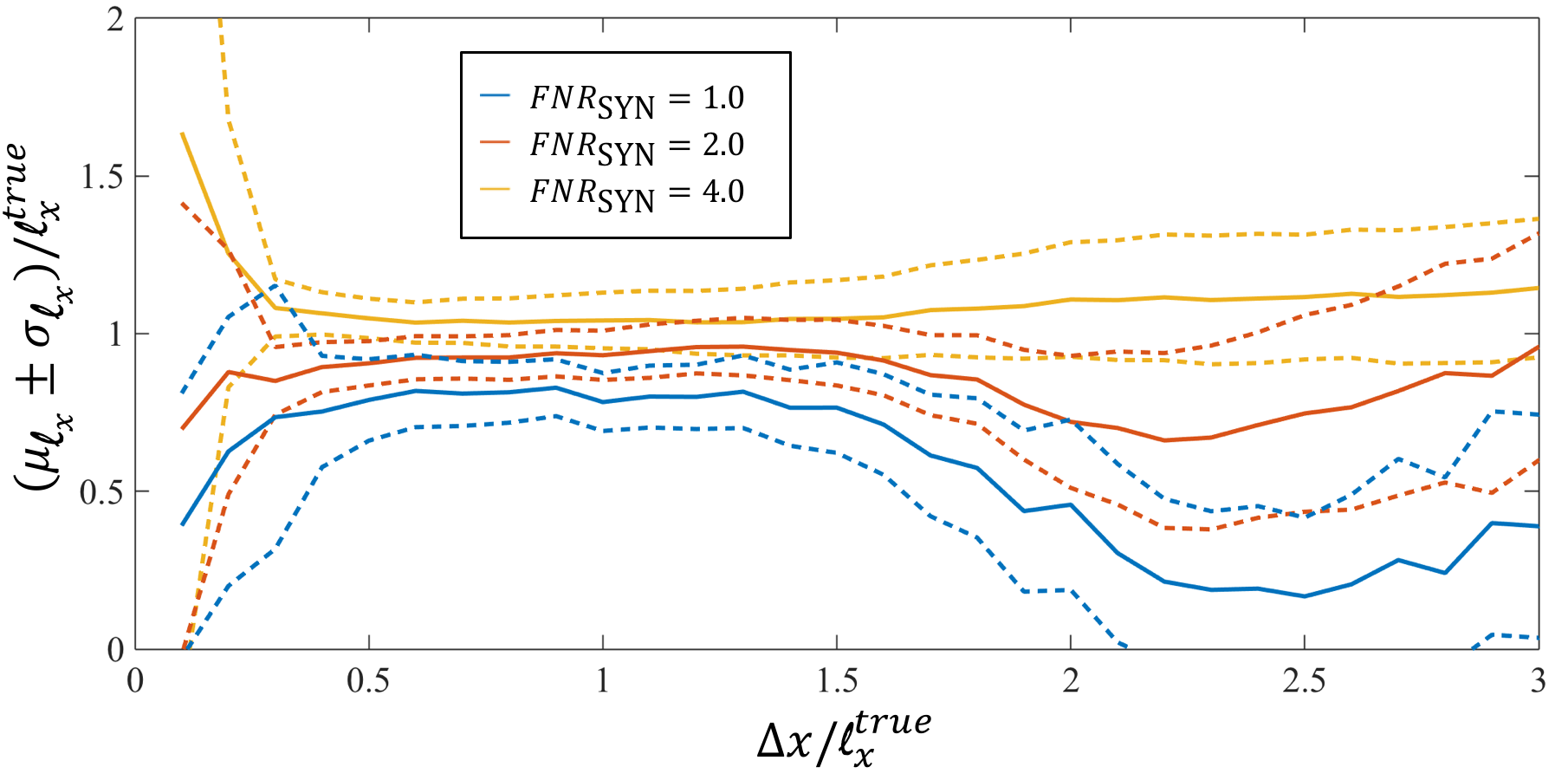}
\caption{Same as \reffig{fig:avg_time_graph_cauchy} for a fixed $\DT = 15,000~\mu$s with the different values of FNR: 1.0 (blue), 2.0 (red) and 4.0 (yellow).}
\label{fig:noise_level_graph_cauchy}
\end{figure}

Once we have the synthetic data with Lorentzian eddies, we follow similar steps as in \refsec{sec:two_point_measurement} with the Lorentzian fitting function:
\begin{equation}
\label{eq:cauchy_fit}
C_2 = C_1 \dfrac{1}{\lp\dfrac{x_a-x_b}{\elxM}\rp^2+1},
\end{equation}
and the measured correlation length ($\elxM$) is estimated to be
\begin{eqnarray}
\label{eq:two_point_fit_cauchy}
\elxM = 
\begin{cases}
\infty & C_1 \le C_2  \nonumber \\
\abs{x_a-x_b}\sqrt{\dfrac{C_2}{C_1-C_2}} & C_1  > C_2  > 0 \\
0 &  C_2 \le 0.
\end{cases} \\
\end{eqnarray}
Note that the covariance function of the Lorentzian function shows that the correlation length ($\elxM$)  is twice as large as the characteristic length ($\lx$), i.e., $\elxT=2\lx$:
\begin{equation}
\label{eq:autocov_cauchy}
\int^{\infty}_{-\infty} \dfrac{1}{\lp\dfrac{x'}{\lx}\rp^2+1} \dfrac{1}{\lp\dfrac{x-x'}{\lx}\rp^2+1} dx'=\dfrac{\pi\lx}{2}\dfrac{1}{\lp\dfrac{x}{2\lx}\rp^2+1}.
\end{equation}

We obtain the expected value $\mu_{\ell_x}$ and the variance $\sigma^2_{\ell_x}$ using \refeq{eq:length_mean_var}. The probability density functions $p\lp C_1\rp=P\lp\lab \overline{C}_{a,a}^\text{noise} \rab\rp$ and $p\lp C_2\rp=P\lp \lab \overline{C}_{a,b}^\text{noise} \rab\rp$ are obtained from the synthetic data, and a Monte-Carlo method is also used to generate histograms similar to \reffig{fig:avg_time_hist} and \reffig{fig:noise_level_hist}. Again, we investigate the accuracy and reliability of the two-point measurement for different sizes of the total time window and different values of FNR.  

To investigate how the size of the total time window affects the measured correlation length, we generate three sets of synthetic data with $\lx = 0.3$ m, $\tlife=15~\mu$s, $v=5,000$ m/s and $\DTsub=585~\mu$s at the fixed FNR of $0.5$, whereas $\DT$ are set to be $10,000~\mu$s, $30,000~\mu$s and $50,000~\mu$s. Note that these values are the same as used to generate \reffig{fig:avg_time_hist}.  \reffig{fig:avg_time_hist_cauchy} shows the histogram of the obtained correlation length with the $p\lp C_1\rp$ and $p\lp C_2\rp$ from the synthetic data, and \reffig{fig:avg_time_graph_cauchy} shows the mean and the standard deviation estimated using \refeq{eq:length_mean_var}. 

For the cases of different values of FNR, we generate the synthetic data with FNR=1.0, 2.0 and 4.0 with $\DT = 15,000~\mu$s; whereas all other values are kept to be same as before. The results are shown in \reffig{fig:noise_level_hist_cauchy} and \reffig{fig:noise_level_graph_cauchy}. We find that the Lorentzian-shaped eddies are not very much different from the Gaussian-shaped eddies, cf. \reffig{fig:avg_time_hist}-\reffig{fig:noise_level_graph}.

\bibliographystyle{apsrev4-1}
\bibliography{2015nf_citation}

\end{document}